\documentclass[AMA,STIXSMALL]{WileyNJD-v2}
\usepackage{moreverb}

\usepackage{amsfonts}
\usepackage{subfig}
\usepackage{graphicx}
\usepackage{fancyref}

  \newcommand{\figref}[1]{\textbf{Figure~\ref{#1}}}
  \newcommand{\Figref}[1]{\textbf{Figure~\ref{#1}}}

  \newcommand{\secref}[1]{Section \ref{#1}}
  \newcommand{\Secref}[1]{Section \ref{#1}}
  \newcommand{\figtext}[1]{{#1}}

\newcommand{\RNum}[1]{\uppercase\expandafter{\romannumeral #1\relax}} 
  \newcommand{\Hanna}[1]{\textcolor{magenta}{(#1)}}

\articletype{Research Article}%

\begin{document}

\title{A Computational Multiscale Model for Contact Line Dynamics}

\author[1]{Hanna Holmgren*}

\author[1]{Gunilla Kreiss}

\authormark{H. Holmgren \textsc{et al}}

\address[1]{\orgdiv{{Department of Information Technology}, \orgname{Uppsala University}, \orgaddress{\state{Box 337, SE-751 05, Uppsala}, \country{Sweden}}}}

\corres{*\email{hanna.holmgren@it.uu.se}}


\abstract[Abstract]{
The conventional no-slip boundary condition leads to a non-integrable stress singularity at a moving contact line. This makes numerical simulations challenging, especially when capillary effects are essential for the dynamics of the flow. This paper presents a new boundary methodology, suitable for numerical simulation of flow of two immiscible and incompressible fluids in the presence of moving contact points. The methodology is based on combining a relation between the apparent contact angle and the contact point velocity with the similarity solution for Stokes flow at a planar interface. The relation between angle and velocity can be determined by theoretical arguments, or from simulations using a more detailed model. The approach here uses the phase field model in a micro domain, with physically relevant parameters for molecular diffusion and interface thickness. The methodology  is used to formulate a new boundary condition for the velocity. Numerical results illustrate the usefulness.
}

\keywords{Two-phase flow, dynamics contact lines}


\maketitle

\section{Introduction}
\label{sec:introduction}
Flow problems involving two immiscible incompressible fluids that are in contact with a solid form so called moving contact line problems. The contact line is located where the interface between the two fluids intersects the solid wall. 

Moving contact line problems form an important class of two-phase flows and appear both in nature and in industrial applications \cite{REW}. Examples include droplet spreading on a solid surface and liquid rising in a narrow tube. Industrial applications where the contact line behaviour is important are coating processes, lubrication, inkjet printing, biological flows and microfluidics such as micropumps and so called lab-on-a-chip devices \cite{ZAHEDI, ROCCA, 1ROCCA, 2ROCCA, 3ROCCA, 4ROCCA,MARTIN}. 

The moving contact line problem has been a subject of debate for many years. The physics governing the dynamics of moving contact lines is still not completely understood \cite{1ROCCA, HUH, REW}. The conventional no-slip boundary condition leads to a non-integrable stress singularity at the contact line \cite{HUH, DUSSAN1}. In fact, molecular dynamics simulations show that there is some sort of fluid slip along the wall in the microscopic region close to the contact line \cite{MOLEC, MOLEC2}. Even if one is interested in the macroscopic behaviour, the dynamics at the contact line is intertwined with the flow at the larger scale in a way that is not possible to model using standard two-phase models \cite{REW}. These small-scale dynamics of the contact line represents a significant numerical difficulty, as it is several orders of magnitude smaller than global flow features in many important applications \cite{MartinMicro}.

\begin{figure}[h!]
  \centering
    \includegraphics [width=0.75\columnwidth]{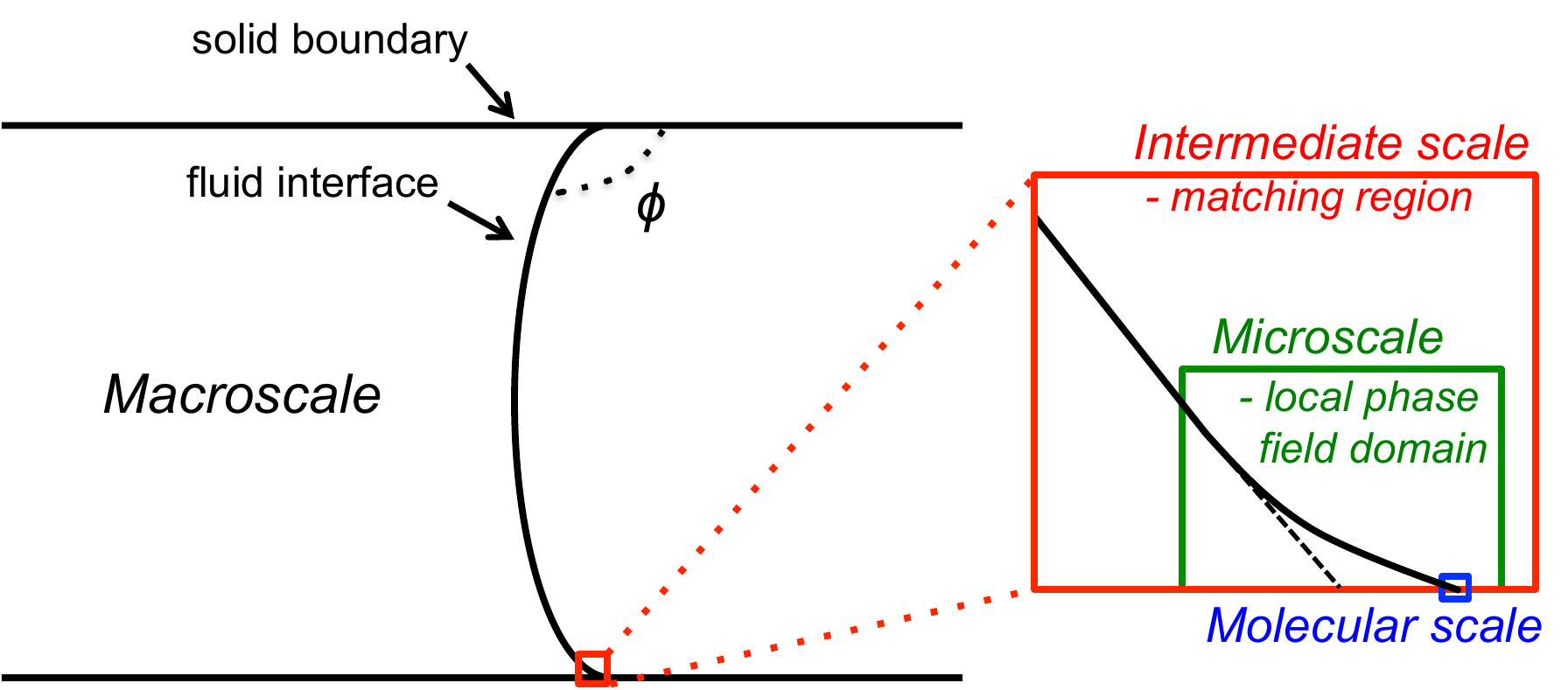}
    \caption{\figtext{Schematic illustration of the different scales.  \label{fig:scales}}}
\end{figure}

The physics at different length scales is illustrated by the successive close-ups near the contact line in \figref{fig:scales}. In the macroscopic region, of typical length scales $L>10^{-7}\,$m
, the flow and the interface shape may be influenced by several different physical phenomena such as capillarity, gravity etc. When capillarity is important, the large scale flow is mainly governed by either inertia and surface tension, or by viscosity and surface tension. Flow situations where inertia and surface tension are dominating are characterized by low Ohnesorge numbers $\mathrm{Oh}=\frac{\mu}{\sqrt{\rho \sigma L}}= \frac{\sqrt{\mathrm{We}}}{\mathrm{Re}} = \sqrt{\frac{\mathrm{Ca}}{\mathrm{Re}}}$, where $\mu$ is the fluid viscosity, $\sigma$ is surface tension, $\rho$ is the fluid density and $L$ is a characteristic length scale \cite{Amberg}. The length scale $L$ could for example be the physical width of a channel, or the  cross section of a droplet or a cavity. Further,  the Weber number, $\mathrm{We}$, relates inertia to surface tension, the Reynolds number, $\mathrm{Re}$, relates inertia to viscous forces and the Capillary number, $\mathrm{Ca}$, relates viscous forces to surface tension.

At smaller scales, illustrated by the intermediate region in \figref{fig:scales}, the flow is governed by a viscocapillary balance \cite{Review2013, CriticalReview}. This balance is characterized by the capillary number $\mathrm{Ca}=\frac{\mu U}{\sigma}$, where $U$ is a characteristic velocity. As mentioned above, the capillary number represents the relative effect of viscous forces versus surface tension acting across an interface. At a viscocapillary balance $Ca\sim 1$.
At this scale the interface shape is influenced by viscous effects and the interface may be strongly curved close to the contact line \cite{DUSSAN3}. Extrapolating the outer solution toward the contact line leads to an apparent macroscopic contact angle $\phi$, see \figref{fig:scales}. 

At the molecular length scale of $L<10^{-9}$ m, the blue region in \figref{fig:scales}, the conventional hydrodynamics breaks down and other models are necessary \cite{Review2013, CriticalReview}. More details on the physics of dynamic contact lines are given for example in the review papers \cite{BONNEGGERS, Review2013, REW}.
One of the best documented methods to overcome the stress singularity at the moving contact line is to replace the no-slip condition with a Navier slip condition and introduce a related slip length parameter \cite{SpeltMacro, Review2013, Legandre}. However, unrealistically large slip length values are necessary for most of these simulations due to grid refinement limitations \cite{Legandre}. 
Additionally, when capillarity is important for the large scale flow, contact line dynamics also need to be accounted for. The slip condition then needs to be combined with a prescribed dynamic contact angle or velocity \cite{REW, SpeltMacro}. The simplest approach is to impose a constant angle corresponding to the static angle $\phi_s$ \cite{RENARDY, KHENNER, LIU2, SPELT_MULTIPLE, AfkhamiBussmann}. 
For cases when the large scale flow is governed mainly by inertia and surface tension (i.e.\ low Ohnesorge numbers), the dynamic angle is close to the static angle \cite{Amberg} and static models can perform rather well \cite{Legandre}.



On the other hand, when the large scale motion is mainly governed by viscosity and surface tension (i.e.\ high Ohnesorge number in capillary driven flows), the evolution is directly dependent on the contact line velocity. For such flows the dynamic or apparent angle is typically different from the static angle \cite{REW, Amberg}, and the contact line velocity is influenced by  small scale features near the contact line. To enhance computational efficiency in large scale simulations some sort of sub-grid modelling can be very beneficial to take such effects into account. A common approach is to prescribe an apparent contact angle according to some empirical law \cite{BONNEGGERS} or hydrodynamics theories \cite{COX, COX2, Hocking, DUSSAN4}. 
In \cite{SpeltMacro, DUPONT, 2ROCCA, Afkhami} 
for example the Cox theory \cite{COX} is used to relate the apparent contact angle to the microscopic angle and contact line speed by
\begin{equation}
g(\phi_L)=g(\phi_s) + \text{Ca}\,\text{log}\left(\frac{L}{\lambda}\right).
\label{eq:cox}
\end{equation}
Here $\lambda$ is the slip length and $L$ is the macroscopic cut-off length scale imposed by the grid resolution where the angle $\phi_L$ is measured \cite{Legandre}. The explicit expressions for $g(\phi)$ are given in for example \cite{SpeltMacro}. The Cox relation is either directly applied \cite{SpeltMacro, DUPONT} 
or by using an adjustable parameter that needs to be empirically determined from experiments \cite{2ROCCA, Afkhami}
. Further, the Cox theory is based on the special case of lubrication theory, and the appropriate dynamic contact angle will depend on the scale on which matching between outer and inner scales occurs \cite{BONNEGGERS}.
A different approach is found in the phase field method, where a  Cahn-Hilliard equation describes the dynamics of the fluid interface, and molecular processes at the interface between fluids and at the contact line are modeled by diffusion \cite{JACQMIN}. In this model, contact line dynamics is handled by a boundary condition relating the surface energy to the contact angle via Young's relation. With this approach there is no stress singularity when the no-slip  boundary condition is used for velocity. However, for accuracy the diffusion processes need to be modeled at physically relevant length scales, which means tens to hundres of nanometers \cite{PHASE-FIELD-SHARP}. This becomes computationally very demanding, and therefore often unphysically large diffusion parameters are used.

Another option is to use a multiscale approach where different physical descriptions at different length scales are coupled using for example the heterogeneous multiscale method \cite{HMM, HMMRev, HMMcomplex}. 
The micro model is usually based on molecular dynamics \cite{MOLEC2, MOLEC5, MOLEC6, REN1}. 
However, these multiscale models have only been applied to two-phase systems in Couette or Poisseuille flows where the densities and viscosities are assumed to be the same in the two fluids. Different contact line models are reviewed in for example \cite{REW, CriticalReview, Legandre}.

In this paper we will describe a multiscale method for simulating contact point dynamics where a conventional macroscale solver is coupled to the local phase field solver in \cite{MartinMicro}. 
There, phase field computations determine a quasi-steady state in a contact point region for a particular apparent contact angle. The relation between the angle and the corresponding contact point velocity characterizes the local dynamics in the contact point region, as discussed in \cite{Review2013}. The phase field simulations in \cite{MartinMicro} are carried out using physically correct diffusion coefficients, and in a domain of corresponding size. The length scale of this domain will here be referred to as the microscopic length scale, and it covers the molecular and parts of the intermediate length scale mentioned earlier. The microscopic length scale is illustrated by the green region in the schematic illustration in \figref{fig:scales}.
As discussed above the phase field model must resolve features at the microscopic scale 
to accurately model the contact point dynamics. A direct coupling between a local domain, modeled by the phase field method, and a macrodomain modeled by the standard Navier--Stokes system would require the macro resolution to match the micro resolution at the interface between the two domains. Due to the scale separation such a matching would be computationally challenging. 

We  take a different approach, where we  use the quasi-steady state description of the local dynamics in the contact point region. A particular apparent contact angle and a particular contact point velocity characterize the dynamics at each moment in time. A main ingredient is  the Huh and Scriven similarity solution \cite{HUH}, which describes the flow around a plane interface at zero Reynolds number, with the interface meeting a solid wall at a well defined contact angle. In many settings the similarity solution is a good approximation of the flow at the intermediate length scale, see \cite{Review2013}.  We will use the similarity solution to bridge the gap between the scales of the micro region and those of the global features of the flow. Based on the Huh and Scriven similarity solution we formulate boundary conditions for the Navier--Stokes system at an artificial boundary, which is placed at a distance from the contact point such that the features of the flow can be resolved at the global scale.

The aim of the paper is to describe the idea of our approach, and to show a possible implementation. We focus on flow situations where the dynamic contact angle differs from the static angle (and methods imposing a static angle are not appropriate). For capillary driven flows, these situations are characterized by high Ohnesorge numbers. Our starting point is a Navier--Stokes solver for two-phase flow coupled to the conservative level set method  described in \cite{CONSLS, CONSLS2}. The level set function enables computing geometric quantities of the interface, such as normal  direction and curvature.  The algorithm  in \cite{CONSLS, CONSLS2} is developed for problems without contact lines. 
We emphasize that in the presence of contact lines, some parts of the methodology may not be as accurate as before. For example mass conservation is no longer guaranteed. 
However, high accuracy for evaluation of geometric quantities and conservation is not in the scope of this work. 

The paper is organized as follows. In \secref{sec:twophasemodel} we start by presenting the macroscopic two-phase model used here. Then we proceed by describing the new multiscale model for dynamic contact points in \secref{sec:contactlinemodel} and implementational details in \secref{sec:implementationofcontactlinemodel}. \Secref{sec:numericalresults} presents numerical simulation results for two test problems and finally a summary and conclusions are given in \secref{sec:conclusions}.

\section{Two-Phase Flow Model}
\label{sec:twophasemodel}
In this section we describe the macroscopic two-phase model used to perform the numerical simulations in \secref{sec:numericalresults}. We also give a brief description of the discretizations of the equations from the two-phase model.

\subsection{Navier--Stokes equations}
\label{sec:NS}
The motion of two immiscible fluids is given by the incompressible Navier--Stokes equations for velocity $\textbf{u}$ and pressure $p$ in non-dimensional form,
\begin{equation}
  \begin{aligned}
  \rho\frac{\partial{\bold u}}{\partial t} + \rho {\bold u} \cdot \nabla {\bold u} &= -
  \nabla p +\frac{1}{\mathrm{Re}}\nabla \cdot (2\mu\nabla^{\mathrm{s}}{\bold u})
  +\mathrm{\textbf{F}}_{st},
  \\
  \nabla \cdot {\bold u}&=0.
\label{eq:NS}
  \end{aligned}
\end{equation}

Here, $\mathrm{\textbf{F}}_{st}$ is the surface tension force at the fluid interface and $\mathrm{Re}$ denotes the Reynolds number, which controls the magnitude of viscous stresses relative to advection. Further, $\nabla^{\mathrm{s}}\bold{u} = \frac{1}{2}(\nabla \bold u + \nabla \bold u^T)$ denotes the rate of deformation tensor and the parameters $\rho$ and $\mu$ denote the density and viscosity measured relative to the parameters of fluid 1,
\[
\rho = \left\{\begin{array}{ll} 1 & \text{in fluid 1,} \\ \frac{\rho_2}{\rho_1} & \text{in fluid 2,} \end{array} \right. \qquad \mu = \left\{\begin{array}{ll} 1 & \text{in fluid 1,} \\ \frac{\mu_2}{\mu_1} & \text{in fluid 2.} \end{array} \right.
\]

\subsection{The conservative level set method}
\label{sec:LS}
The level set method is used to keep track of the fluid-fluid interface and the moving contact point. The level set function $\varphi(\textbf{x}, t)$ is used as an indicator function, where the fluid-fluid interface $\Gamma$ is given by the zero level set of $\varphi$. The subdomain $\Omega_1$ occupied by fluid 1 is given by $\varphi > 0$ and the subdomain $\Omega_2$ occupied by fluid 2 is given by $\varphi < 0$. The interface is here captured by the conservative level set method developed in \cite{CONSLS, CONSLS2}. The indicator function $\varphi$ is a smoothed color function; the function smoothly switches value form $+1$ to $-1$ in a transition region around the interface. At the initial time, $\varphi$ is computed from a signed distance function $d({\bold x})$ around the interface by 
  \begin{equation}
  \varphi(\cdot,0)=\tanh\left({\frac{d({\bold x})}{\varepsilon}}\right),
  \label{eq:color}
  \end{equation}
where $\varepsilon$ is a parameter that controls the thickness of the transition region.
The level set function is advected in time by the underlying fluid velocity according to the Hamilton--Jacobi equation
\begin{equation}
\frac{\partial \varphi}{\partial t}+  {\bold u} \cdot \nabla \varphi= 0.
\label{eq:LS}
\end{equation}

After advecting the fluid interface, the surface tension force $\mathrm{\textbf{F}}_{st}$ is calculated using the approximation of a smeared surface tension according to \cite{zahedispurious},
\begin{equation}
 \mathrm{\textbf{F}}_{st}= \frac{1}{\mathrm{We}}\kappa \nabla H_{\varphi}, 
 \label{eq:surface_tension}
\end{equation}
where $\kappa$ is the interface curvature, $\mathrm{We}$ is the Weber number and $H_{\varphi}$ is here given by
\begin{equation}\label{eq:heaviside}
H_\varphi \equiv \mathbb{H}_\varepsilon (d({\bold x})), \quad d({\bold x}) = \log\left(\frac{1+\varphi({\bold x})}{1-\varphi({\bold x})}\right). 
\end{equation}
The signed distance function $d({\bold x})$ is reconstructed from $\varphi$. The function $\mathbb{H}_\varepsilon$ denotes a one-dimensional smoothed Heaviside function that changes value from $0$ to $1$ over a length scale proportional to $\varepsilon$. The interface curvature $\kappa$ can be computed using the level set function by
\begin{align}
& \kappa=-\nabla \cdot {\bold n} , \label{eq:normcurv1} \\
\text{where}~\,&{\bold n}=\frac{\nabla \varphi}{|\nabla \varphi|} \label{eq:normcurv2}.  
\end{align}
More details about the surface tension term is given in \cite{zahedispurious} and \cite{MartinHPC}. 

Over time the level set function will loose its shape and thus its relation to the signed distance function due to discretization errors and non-uniform velocity fields. To retain the shape of the level set function, $\varphi$ has to be reinitialized with regular intervals. For the conservative level set method, this is done by solving the following equation to quasi-steady state
\begin{equation}
\frac{\partial \varphi}{\partial \tau}+\nabla \cdot ({\bold n}(1-\varphi^2))-\nabla \cdot ({\bold n}  \varepsilon \nabla \varphi \cdot {\bold n})=0,
\label{eq:reinit}
\end{equation}
where $\tau$ is a pseudo time step and ${\bold n}$ is the interface normal. 
This equation calculates a smoothed color function by balancing diffusion in direction normal to the interface by a compressive flux. If the fluid-fluid interface does not intersect the computational boundary homogenous Neumann boundary conditions can be used for the reinitialization. However, for the case when contact points are present the contact point position must not be distorted during reinitialization and we need other conditions. We discuss this further in \secref{sec:reinit}.


\subsection{Discretizations}
\label{sec:discimp}
For the implementation we use the existing two-phase flow solver described in \cite{MartinHPC} with suitable modifications to account for moving contact points (see \secref{sec:implementationofcontactlinemodel}). 
The solver is implemented in the C++ based finite element open source library deal.ii \cite{DEAL1, DEAL2}. The equations in \secref{sec:NS} and \secref{sec:LS} are discretized in space using the finite element method. For the level set function piecwise continuous linear shape functions on quadrilaterals, i.e.\ $Q_1$ elements, are used. For the incompressible Navier--Stokes equations we use the Taylor--Hood elements $Q_2Q_1$, i.e.\ shape functions of degree two for each component of the velocity and of degree one for the pressure. With these elements the Babu\^ska--Brezzi (inf--sup) condition \cite{INFSUP} is fulfilled in order to guarantee the existence of a solution. 

Finite element discretizations of equations of transport type, such as the level set equation, typically need to be stabilized. Here however, no stabilization is used since the reinitialization will take care of possible oscillations. To improve robustness, the equation for curvature calculation \eqref{eq:normcurv1} is solved by projecting the divergence of the normal vector to the space of continuous finite elements with a mesh-dependent diffusion $4h^2$ \cite{MartinHPC, zahedispurious}. 


For time stepping, each of the Navier--Stokes equations and the level set equation are discretized using the second order accurate, implicit BDF--2 scheme. In order to avoid an expensive coupling between the incompressible Navier--Stokes part and the level set part (via the variables $\textbf{u}$ and $\varphi$) a temporal splitting scheme is introduced. For more details about the time discritization we refer to \cite{MartinHPC}.
\section{Contact Line Model}
\label{sec:contactlinemodel}


In this section we 
derive macroscopic boundary conditions for simulations of dynamic contact points. Similarly to the work in \cite{MartinMicro} we assume both a temporal and a spatial scale separation between the local dynamics at the contact point and the global fluid flow. 
For capillary driven flows (or other flows where capillarity is essential) the dynamics at contact points influence the global flow \cite{MartinMicro, Amberg}. For these type of problems the macroscopic time scale is typically large compared to the time it takes for the microscopic problem to reach a steady state \cite{MartinMicro}, why it possible to assume a temporal scale separation. Consequently the microscopic dynamics is in equilibrium for each apparent contact angle and no additional information from the macro model is required \cite{MartinMicro}. This assumption is also supported for example in \cite{DUSSAN3} where it is concluded the macroscopic dynamics is affected by the microscopic regime mainly through the contact angle, and in the review paper \cite{Review2013} where they note that in many flow situations where Ca is small the apparent contact angle completely describes the dynamics.


With the scale separations as a motivation, we consider multiscale modelling, where it is possible to use different models for describing the microscopic, mesoscopic (intermediate) and macroscopic dynamics. 
The coupling of the models is done via the macroscopic contact angle and the local contact point velocity. 
Here, we focus on how to communicate the information about the local contact point velocity from the microsimulation to the macro simulation via macroscopic boundary conditions. 

The macroscopic moving contact point model and boundary conditions we develop here will directly make use of the local contact point model developed in \cite{MartinMicro}. The model in \cite{MartinMicro} describes the dynamics at a length scale of tens to hundreds of nanometers, i.e.\ the green region in the schematic illustration from \figref{fig:scales}. The idea is based on performing a series of simulations using the Cahn--Hilliard and Stokes equations for the micro scale dynamics, and taking the molecular effects into account by using a standard phase field boundary condition. The model takes the apparent contact angle as input and calculates the local contact point velocity. The viscous bending of the interface is to a large extent captured by this model.  
Since the contact point velocity only depends on the apparent contact angle (under the assumptions made here) it is possible and most efficient to perform the local simulations in \cite{MartinMicro} independently of the macro simulation. The results from the local phase field simulations can then be tabulated and efficiently employed by the macromodel. We refer to \cite{MartinMicro} for more details about the local model.

It is possible to couple the macroscopic boundary conditions developed here to any local simulation that calculates the local contact point velocity from a given apparent contact angle. Another option would be to couple the macroscopic boundary conditions to a molecular dynamics simulation. However, since molecular dynamics simulations are limited to lengths scales of $L<10^{-9}$ (only the blue region in \figref{fig:scales}) an extra mesoscopic model would be necessary for the intermediate mesoscale dynamics.
	

\subsection{Matching} 
\label{sec:matching}
To couple the local phase field model and the macroscopic model we use ideas from matched asymptotics. Just as in \cite{MartinMicro} the macroscopic scale, with an apparent contact angle, represents an outer solution and the microscopic scale represents an inner solution.  The matching between the outer and inner solutions is done at the intermediate mesoscale, close to the contact point at the outer scale but far from the contact point at the inner scale \cite{MartinMicro}. The continuum approximation is assumed to be valid in the intermediate region, but the length scale is small enough for  viscous effects to dominate the convection. By this assumption we have a vanishing local Reynolds number (based on the characteristic length scale of the intermediate region) and  the creeping flow approximation of the Navier--Stokes equations can be employed.

Asymptotically the interface at the microscale becomes increasingly planar far from the contact point \cite{ MartinMicro}. This is seen theoretically where a logarithmic dependence of the contact angle is predicted \cite{BONNEGGERS, Review2013}: the curvature of the asymptotic microscopic solution goes to zero as the distance to the contact point goes to infinity. From the logarithmic dependence it can also be concluded that viscous bending of the interface is most prominent near the contact point and will to a large extent be captured by the phase field model in the microdomain.  Consequently, the contact angle varies only very slowly at the matching scale and we approximate the interface to be essentially planar at this scale \cite{MartinMicro}. 

The flow around a flat fluid interface under the assumption of a creeping flow approximation (wedge flow) was studied theoretically by Huh and Scriven \cite{HUH}. In their paper Huh and Scriven derive an analytic similarity solution to the creeping flow approximation of Navier--Stokes equations by rewriting them in the form of a biharmonic equation for the stream function $\psi(r, \theta)$ (in plane polar coordinates $r$ and $\theta$). The origin of the polar coordinate system is fixed to the contact point position. In terms of the stream function the polar velocity components are $v_r = -r^{-1}\frac{\partial \psi}{\partial \theta}$ and $v_{\theta} = \frac{\partial \psi}{\partial r}$. By imposing appropriate boundary and interface conditions an analytical expression for the stream function in the region close to the contact point is derived. The analytical Huh and Scriven solution is a function of the contact angle $0 < \phi < 180$, the magnitude of the contact point velocity $U$ and the viscosity ratio $Q$. It is given by
\begin{equation}
\psi(r,\theta)= r(a \sin\theta+b\cos\theta+c\theta\sin\theta+d\theta\cos\theta ), 
\end{equation}
where the coefficients for the two different fluids are given by (subscripts denotes fluid 1 and 2 respectively)
 \begin{align}
 &a_1=-U-\pi c_1-d_1  \\
 &b_1=-\pi d_1 \\
 &c_1=US^2[S^2-\gamma\phi+Q(\phi^2-S^2)]/D    \\
 &d_1=USC[S^2-\gamma\phi+Q(\phi^2-S^2)-\pi \tan\phi]/D \\
 &a_2=-U-d_2  \\
 &b_2=0 \\
 &c_2=US^2[S^2-\gamma^2+Q(\delta\phi-S^2)]/D  \\
 &d_2=USC[S^2-\gamma^2+Q(\delta\phi-S^2)-Q\pi \tan \phi]/D  ,
 \end{align}
 where 
\begin{align}
&S=\sin\theta \notag \\
&C=\cos \theta \notag \\
&\gamma=\phi-\pi \notag \\
&Q=\mu_A/\mu_B \notag \\
&D=(SC-\phi)(\gamma^2-S^2)+Q(\delta-SC)(\phi^2-S^2) .\notag 
\end{align}

\begin{figure}[h!]
  \centering
    \includegraphics [width=0.7\columnwidth]{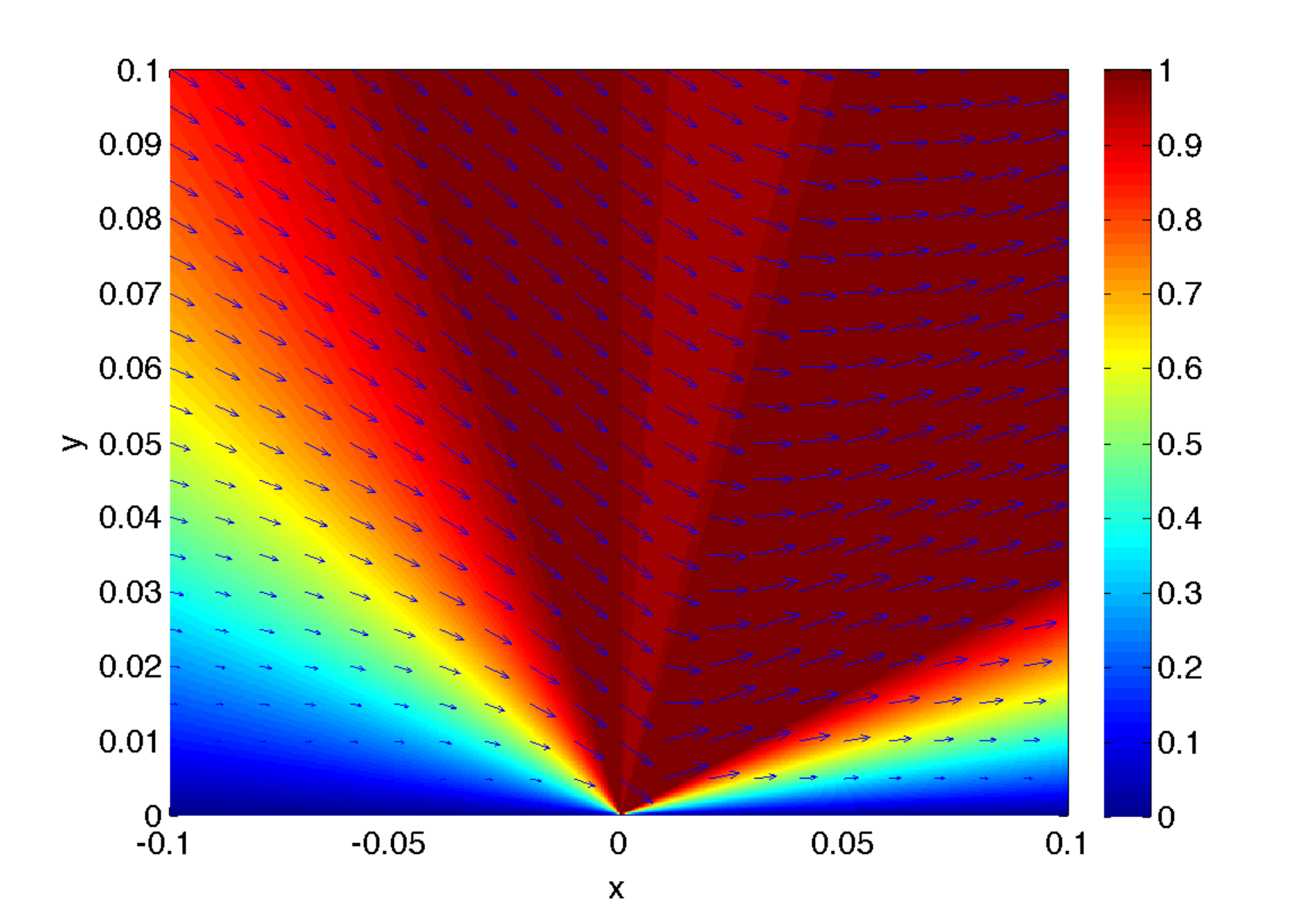}
    \caption{\figtext{The analytical velocity field in the intermediate region for $\phi = 45$, $U = 1$ and $Q=1$.}}
    \label{fig:simsol}
\end{figure}

In \figref{fig:simsol} the magnitude of the Huh and Scriven similarity velocity is plotted for the case with a contact angle $\phi = 45$, non-dimensional contact velocity $U = 1$ and viscosity ratio $Q = 1$.
It can be seen that the velocity is zero along the whole solid boundary, i.e.\ along the whole line $y = 0$, also at the contact point. However, just inside the boundary (i.e.\ along the line $y=\epsilon$, where $\epsilon$ is a small number) the velocity is non-zero in the vicinity of the contact point. This illustrates a velocity discontinuity at the contact point.
The Huh and Scriven solution is recognised to be a useful tool to describe flow at an intermediate scale close to contact points, see for example the review paper \cite{Review2013}, and also \cite{Yulii,DUSSAN,DUSSAN1}. It is also well known that this solution is not a full solution and cannot be used to describe the full moving contact point problem. The solution is singular exactly at the contact point. However, in this small region atomistic phenomena come into play and are not included in the standard Navier--Stokes system. If the Huh and Scriven solution were regular there, it would not be physically relevant. Further, a completely planar interface is unrealistic. There is a jump in the pressure over the interface, which can only be balanced by surface tension of a curved surface. However, if the surface tension effect is strong, which is the case for flow driven by capillary forces, a small curvature suffices.  In \cite{chen1997} a modified Huh and Scriven solution is presented, which reveals that at an intermediate length scale the curvature decreases away from the contact point, and the flow approaches the flow of the planar case. In the paper \cite{MartinMicro} results from numerical simulations using the phase field solution also shows agreement with the wedge flow close to the contact point (at the macroscopic scale). Furthermore, Huh and Scriven make a similar conclusion in their paper: "the model may approximate reality well in a slightly removed region where the fluid interface is substantially flat and the flow qualifies as creeping" \cite{HUH}.
The conclusion is that the planar interface and the Huh and Scriven solution are appropriate for matching an outer solution with an inner solution, at some intermediate distance from the contact point.

\subsection{Macroscopic boundary conditions}
\label{sec:macroscopicboundaryconditions}
\begin{figure}[h!]
  \centering
    \includegraphics [width=0.45\columnwidth]{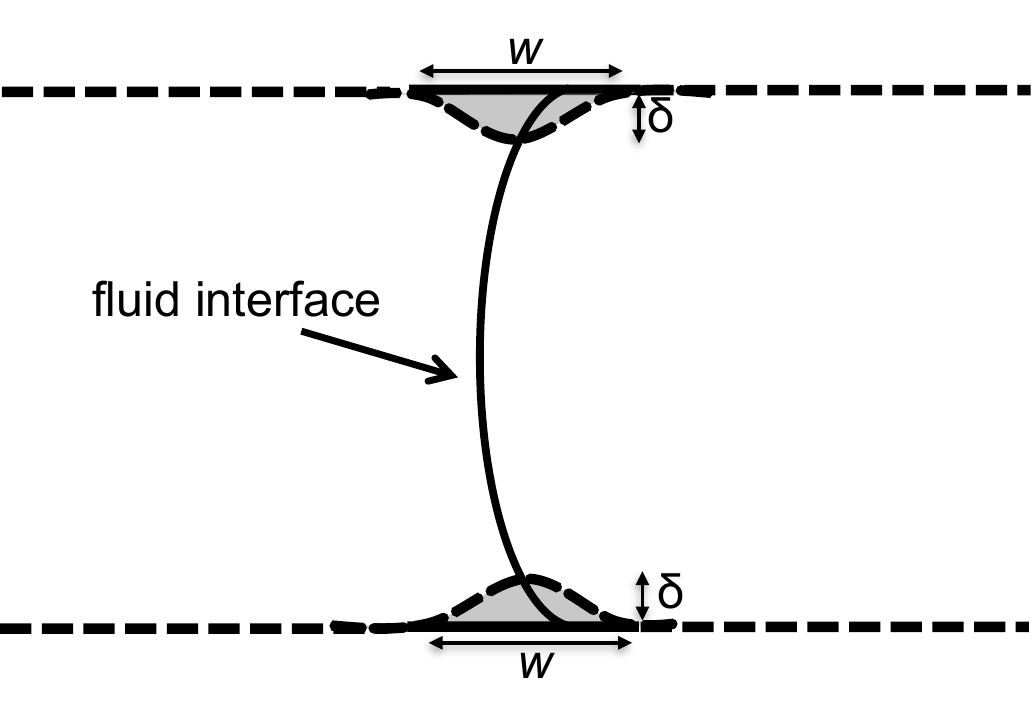}
    \caption{\figtext{Schematic illustration of the modified computational domain with an artificial boundary (dashed) where the macroscopic boundary conditions are applied. }}
    \label{fig:mod_dom}
\end{figure}

With the motivation from previous subsection we use the analytic Huh and Scriven solution to develop the macroscopic velocity boundary conditions. To avoid the singularity at the contact point in the analytical solution, parts of the intermediate matching region is excluded from the macroscopic simulation. If the excluded region is of height $\delta$ and width $w$ this will result in a modified boundary according to the schematic illustration in \figref{fig:mod_dom}. In the excluded region the flow is given by the local model from \cite{MartinMicro} and the analytic Huh and Scriven solution. The length scale of the excluded region should be in the order of the length scale of the intermediate region discussed in previous section. Hence, $\delta$ and $w$ are parameters that should depend on the physics of the two fluids.  They need to be small compared to macroscopic features, but large enough compared to molecular diffusion lengths.

\begin{figure}[h!]
  \centering
    \includegraphics [width=0.7\columnwidth]{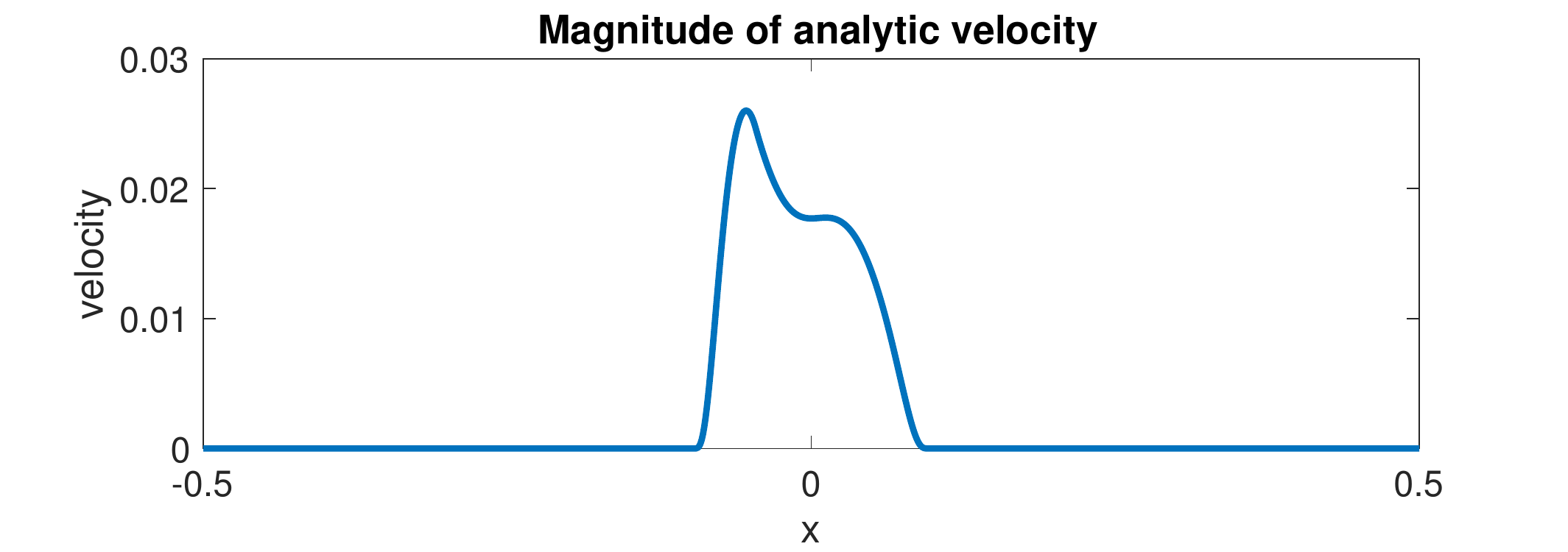}
    \caption{\figtext{The magnitude of the analytical velocity along the artificial boundary in \figref{fig:mod_dom} for an excluded region of size $\delta=0.05$ and $w=4\delta$, a given non-dimensional contact point velocity $U=0.02$ and contact angle $\phi=140$. }}
    \label{fig:fullline}
\end{figure}

Along the new artificial boundary, we impose the analytic velocity from the Huh and Scriven model as a velocity Dirichlet boundary condition for the macroscopic simulation. In this way the information about the contact point velocity $U$ from the micro model, i.e.\ the information about the movement of one single point, is transformed into a macroscopic velocity boundary condition along the modified boundary. \Figref{fig:fullline} shows an example of such a boundary function. The magnitude of the analytic Huh and Scriven velocity along the artificial boundary is plotted for the case when a region of size $\delta=0.05$ and $w=4\delta$ has been excluded, for a non-dimensional contact point velocity $U=0.02$ and contact angle $\phi=140$. Note that the relation between contact angle and contact point velocity involves a non-dimensional velocity obtained from the local phase field model. To get the corresponding physical velocity we need to take  the scaling by the reference velocity $U_{ref}=\sigma/\mu$ into account, see \cite{MartinMicro} for more details.

Further, we also note that in our model the shape of the fluid interface is not prescribed at the boundary. The level set function develops dynamically as part of the solution and the contact angle is calculated from the resulting level set function.
The temporal evolution of the level set function is modeled by the advection equation, and the parabolic reinitialization equation. For the reinitialization equation, Dirichlet boundary conditions are imposed, while no boundary condition is imposed in the advection step, see subsequent section. We use this option, since numerical experiments have shown that prescribing Dirichlet or Neuman boundary conditions in the advection step distorts the contact angle.

\section{Implementation of Contact Line Model}
\label{sec:implementationofcontactlinemodel}
In this section we describe  implementational details related to the contact point model presented in previous section.
\subsection{Artificial boundary} 
\label{sec:artificialboundary}
The modified boundary (\figref{fig:mod_dom}) is described using a so called bump function $f(x)$:
\begin{align}
 f(x)= 
\begin{cases}
\delta \, e^{\left(1-{{\left[1-\left({\frac{x-x_{cp}}{w/2}}\right)^2\right]}^{-1}}\right)}& \text{if } |x-x_{cp}|<\frac{w}{2}\\
    0              & \text{otherwise},
\end{cases}
\end{align}
where $\delta$ and $w$ is the height and width of the bump function respectively (see \figref{fig:mod_dom}) and $x_{cp}$ is the $x$-coordinate of the contact point position. 
A fist simple approach to implement the macroscopic boundary conditions is to  apply the velocity boundary condition at the gird points along the physical boundary, but read the values of the analytical velocities from the modified boundary (i.e.\ from the bump in \figref{fig:mod_dom}). In this work we are interested in a first investigation to see if the contact point model is able to advect the contact point accurately and we will perform simulations of model problems. With this motivation, we start with this simple approach of directly projecting the values of the velocity boundary function along the modified boundary to the physical boundary.

\begin{figure}[h!]
  \centering 
 \includegraphics [width=0.75\columnwidth]{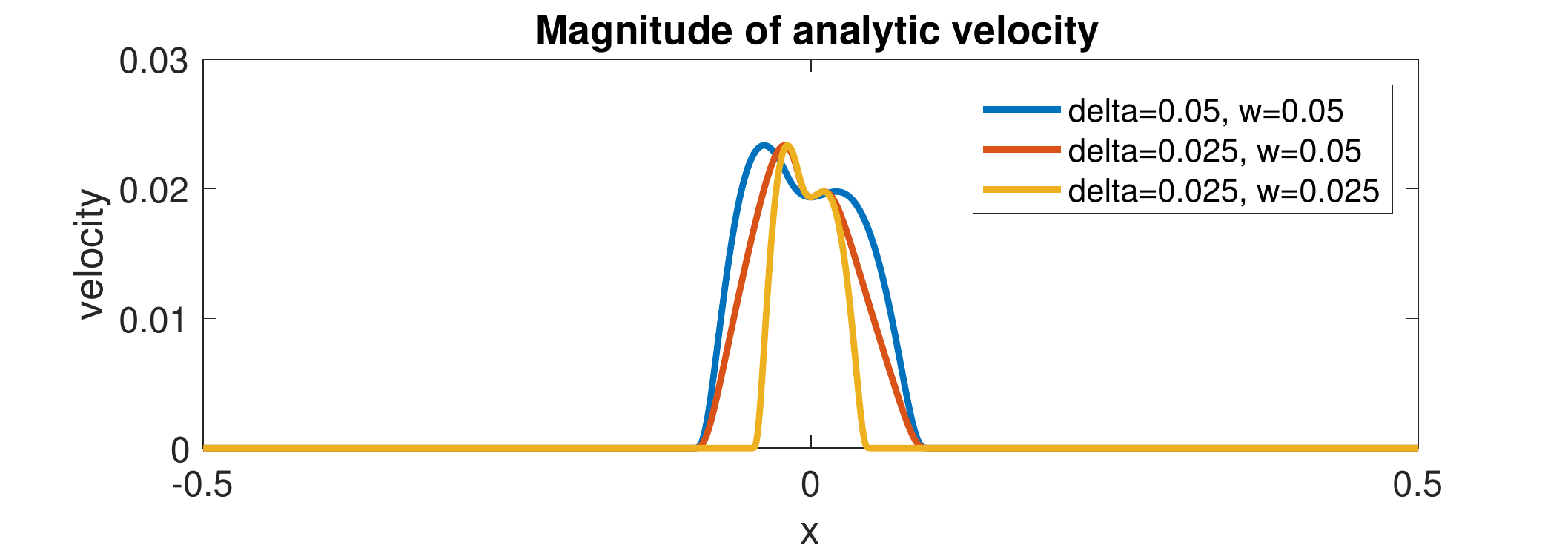}
    \caption{\figtext{The features of the velocity boundary function need to be resolved by the mesh. Here we plot an example of a boundary function for different values of $\delta$ and $w$ ($U=0.02$ and $\phi=112$ for all plots). }}
    \label{fig:delta_dep}
\end{figure}

The size of the excluded domain, i.e.\ $\delta$ and $w$, puts a restriction on the spatial mesh size. It is important to resolve the small scale features of the peak in the velocity boundary function around the contact point, see \figref{fig:fullline} for example. \Figref{fig:delta_dep} shows an example of how the peak and the small scale features of the boundary function depends on $\delta$ and $w$. 


\subsection{Calculation of contact point position and angle}
\label{sec:angle}
\begin{figure}[h]
\centering
  \includegraphics[width=0.5\columnwidth]{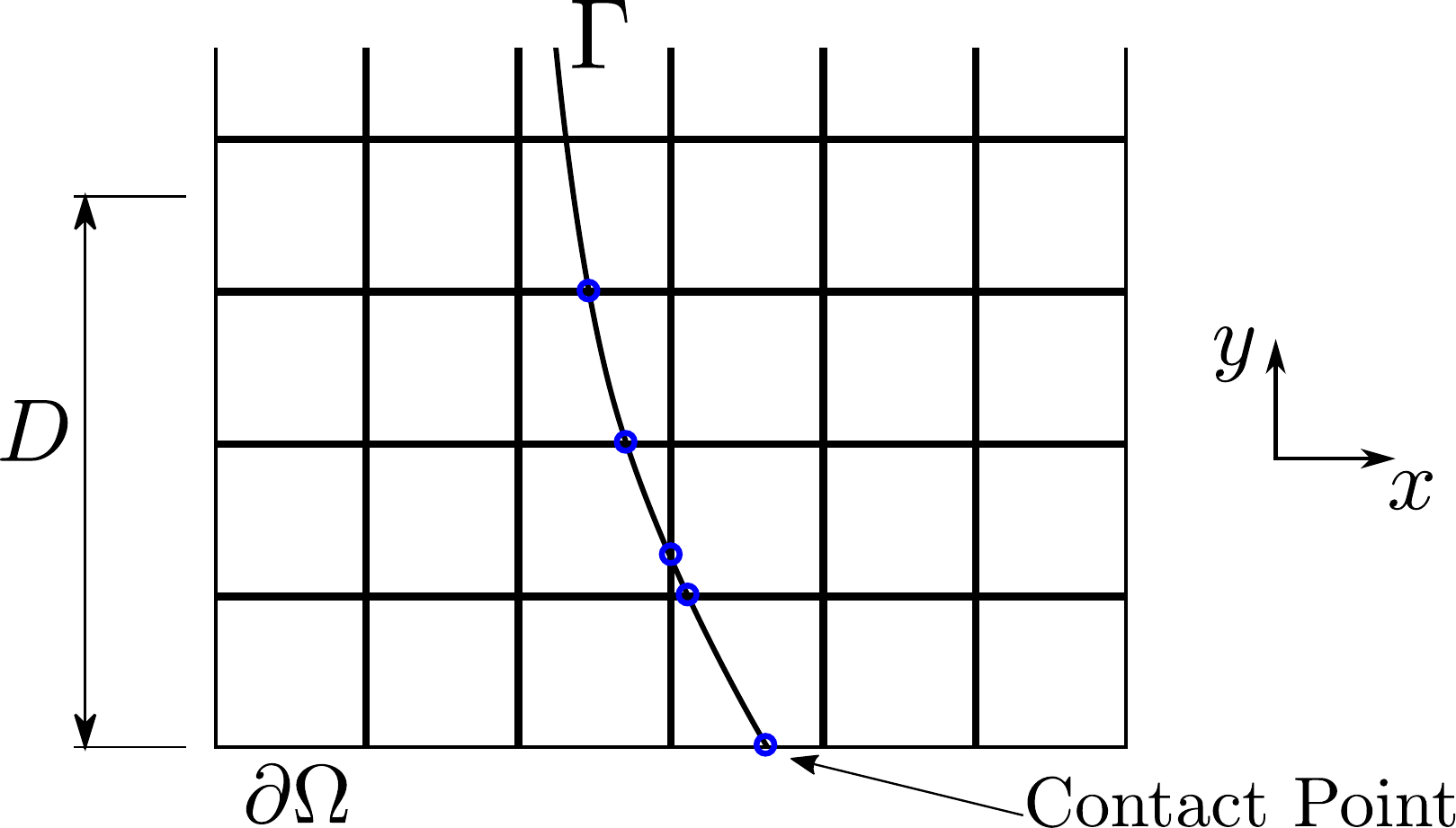}
  \caption{Intersections between the mesh faces and the interface. \label{fig:pointAndAngleDetection}}
\end{figure}

In order to calculate the contact point and contact angle, the method illustrated in \figref{fig:pointAndAngleDetection} is used.
First, all intersections points, $(x_i,y_i)$, between the fluid-fluid interface $\Gamma$ and the mesh faces are found, that are closer to the boundary than a distance $D$.
Among these points, the point located on the domain boundary, $\partial \Omega$, is taken as the contact point.
The interface is then approximated by a second order polynomial $x=c_1+c_2y+c_3y^2$ by making a 
least square fit in order to minimize 
\begin{equation*}
\sum_i \|c_1+c_2 y_i + c_3y^2 - x_i \|^2.
\end{equation*}
The contact angle is then computed from the slope of the second order polynomial
\begin{equation*}
\frac{dx}{dy}=c_2+2c_3y,
\end{equation*}
evaluated at the contact point position $y=y_{cp}$.

\subsection{Reinitialization at boundaries with contact points}
\label{sec:reinit}
As discussed in \secref{sec:LS} we use the conservative level set reinitialization according to equation \eqref{eq:reinit}. For the case when contact points are present, using the standard homogenous Neumann boundary condition for reinitialization will distort the contact point position. To fix the contact point position during the reinitialization we instead use a Dirichlet boundary condition at the boundaries with contact points. To minimize the distortion of the contact angle we base the Dirichlet boundary condition on a contact angle that is calculated before the reinitialization cycle. The boundary condition is based on approximating the fluid-fluid interface to be a circular arc, with the calculated contact angle, in an area of size $\sim\varepsilon^2$ close to the contact point (where $\varepsilon$ is the parameter that controls the transition region of the level set function, see \secref{sec:LS}). Thus we force the level set function on the boundary to take the form
\begin{equation}
\label{eq:reinitBC}
\phi = \text{tanh}\bigg(\frac{d(\textbf{x})}{\varepsilon}\bigg) \text{~ on } \partial \Omega,
\end{equation}
were $d(\textbf{x})$ is the signed distance function around the interface according to a circular arc interface. Note that this boundary condition will not force the fluid-fluid interface to take the form of a circular arc inside the domain, it will only effect the level set function on the boundary. 
Also, since the color function \eqref{eq:reinitBC} takes the value $+1$/$-1$ along the boundary except in a region close to the contact point ($x_{cp}\pm \varepsilon$), it is only a local approximation. 

\subsection{Summary of full solution algorithm}
To summarise, we look at the full solution algorithm. 
In each time step of the simulation we perform the following steps:
\begin{enumerate}
	\item	Calculate the contact point position $x_{cp}$ and the apparent contact angle $\phi$ according to \secref{sec:angle}.
	\item For the apparent contact angle $\phi$, obtain the local contact point velocity $U$ from the pre-tabulated data from the local phase field model in \cite{MartinMicro}. 
	\item Set up the macroscopic boundary conditions according to \secref{sec:macroscopicboundaryconditions}. These depend on the local contact point velocity $U$, the contact point position $x_{cp}$ and the calculated contact angle $\phi$. 
	\item Calculate the interface curvature and normals according to equations \eqref{eq:normcurv1}--\eqref{eq:normcurv2}.
	\item	Compute the surface tension force in \eqref{eq:surface_tension}.
	\item	Solve the Navier--Stokes equations \eqref{eq:NS} with the macroscopic velocity boundary conditions and the calculated surface tension force.
	\item	Advect the level set function according to equation \eqref{eq:LS} with the underlying fluid velocity field from the Navier--Stokes solution.
	\item	Every third time step:
	\vspace{-0.25cm}
	\begin{itemize}
		\item	Calculate the contact point position $x_{cp}$ and apparent contact angle $\phi$ again.
		\item	Reinitialize the level set function according to equation \eqref{eq:reinit} (three pseudo time steps), with the boundary condition in \secref{sec:reinit} at boundaries with contact points.
	\end{itemize}
\end{enumerate}

\section{Numerical Experiments and Results}
\label{sec:numericalresults}

\begin{figure}[h!]
  \centering
    \includegraphics [width=0.6\columnwidth]{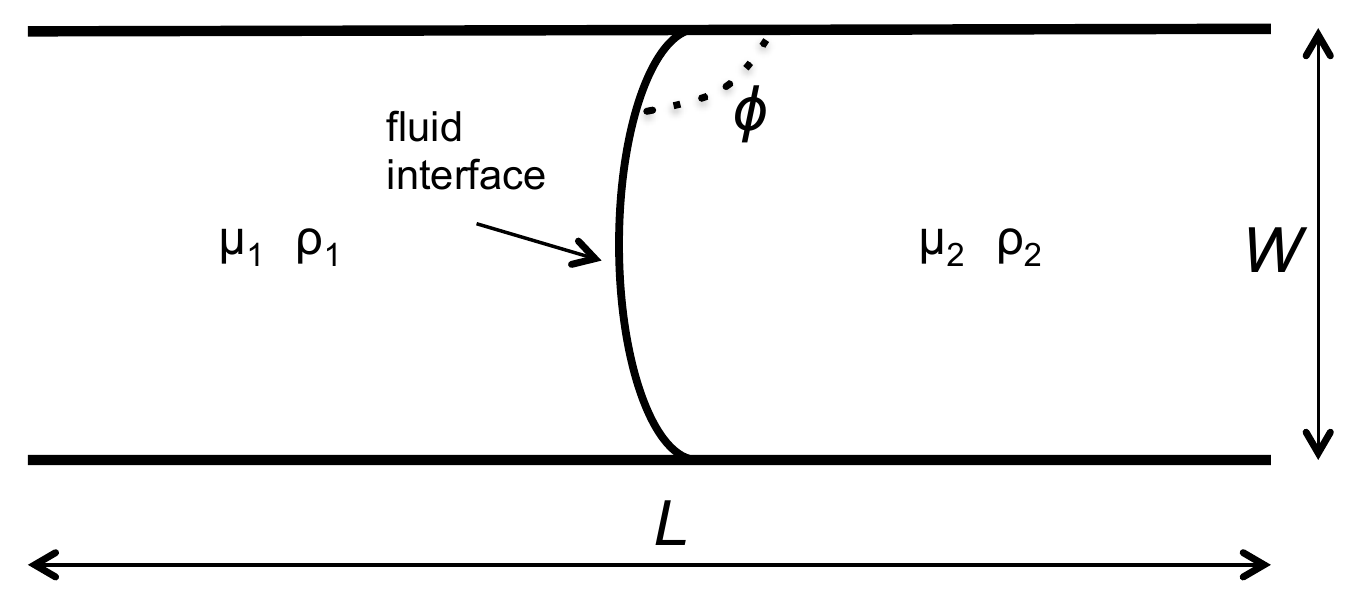} 
    \caption{\figtext{Model problem: Capillary fluid-fluid displacement in a horizontal channel of length $L$ and width $W$. }}
    \label{fig:setup}
\end{figure}

As a model problem we consider two-dimensional capillary driven flow in a horizontal channel (no gravity). The model domain is illustrated in \figref{fig:setup}. 
We refer to the fluid to the left of the interface as fluid $1$ and the fluid to the right as fluid $2$, and all contact angles are measured from fluid $2$. For all simulations the pressure is fixed to zero at the open boundaries (left and right boundaries). We first present simulation results for a simplified problem (Preliminary Test), and then proceed by demonstrating results for the full capillary driven channel flow (Channel Flow). In both cases we solve the non-dimensional Navier--Stokes equations (\ref{eq:NS}) with $\mathrm{Re}=1$ and surface tension force given by (\ref{eq:surface_tension}), where  $\mathrm{We}=1$.

\subsection{Preliminary Test}
\label{sec:Test1}
To investigate if and how well the macroscopic boundary condition is able to transport the contact point according to a given contact point velocity, we first perform simulations in a simplified set-up. A non-dimensional contact point velocity of $U=0.02$ is prescribed for all contact angles, i.e.\ we assume the micro model gives $U=0.02$ for all contact angles. A channel of non-dimensional length $L=10$ and width $W=2$ is used. For this test, the viscosities and densities are equal in the two fluids, i.e.\ $\mu_1=1$, $\frac{\mu_1}{\mu_2}=1$, $\rho_1=1$, $\frac{\rho_1}{\rho_2}=1$ and we assume a non-dimensional surface tension of $\sigma=2$. The fluid interface is initialized to a circular arc with contact angle $\phi=100$ and the initial velocity is zero in the whole domain. The velocity boundary conditions constructed in \secref{sec:macroscopicboundaryconditions} are applied along the upper and lower wall of the channel. The conditions are constructed according to the prescribed contact point velocity of $U=0.02$ and the estimated apparent contact angles. The apparent contact angles are calculated at a height of $D=0.1$ for all meshes (i.e.\ at the height of two coarse grid cells above the wall, see \secref{sec:angle}).

The simulations are run for a total non-dimensional time of $T=20$ and four different meshes are used. The coarse mesh consists of $40\times200$ grid cells (mesh size $h=0.05$) and the other meshes are uniformly refined two, three and four times respectively. The height and width of the bump function from \secref{sec:artificialboundary} is here $\delta=0.05$ and $w=4 \delta$ (i.e.\ $\delta=0.025W$). A time step of $\Delta t=0.2$ is used for the coarse mesh, and then reduced to keep the CFL number constant when the mesh is refined.  

\begin{figure}[h!]
  \centering
    \includegraphics [width=0.5\columnwidth]{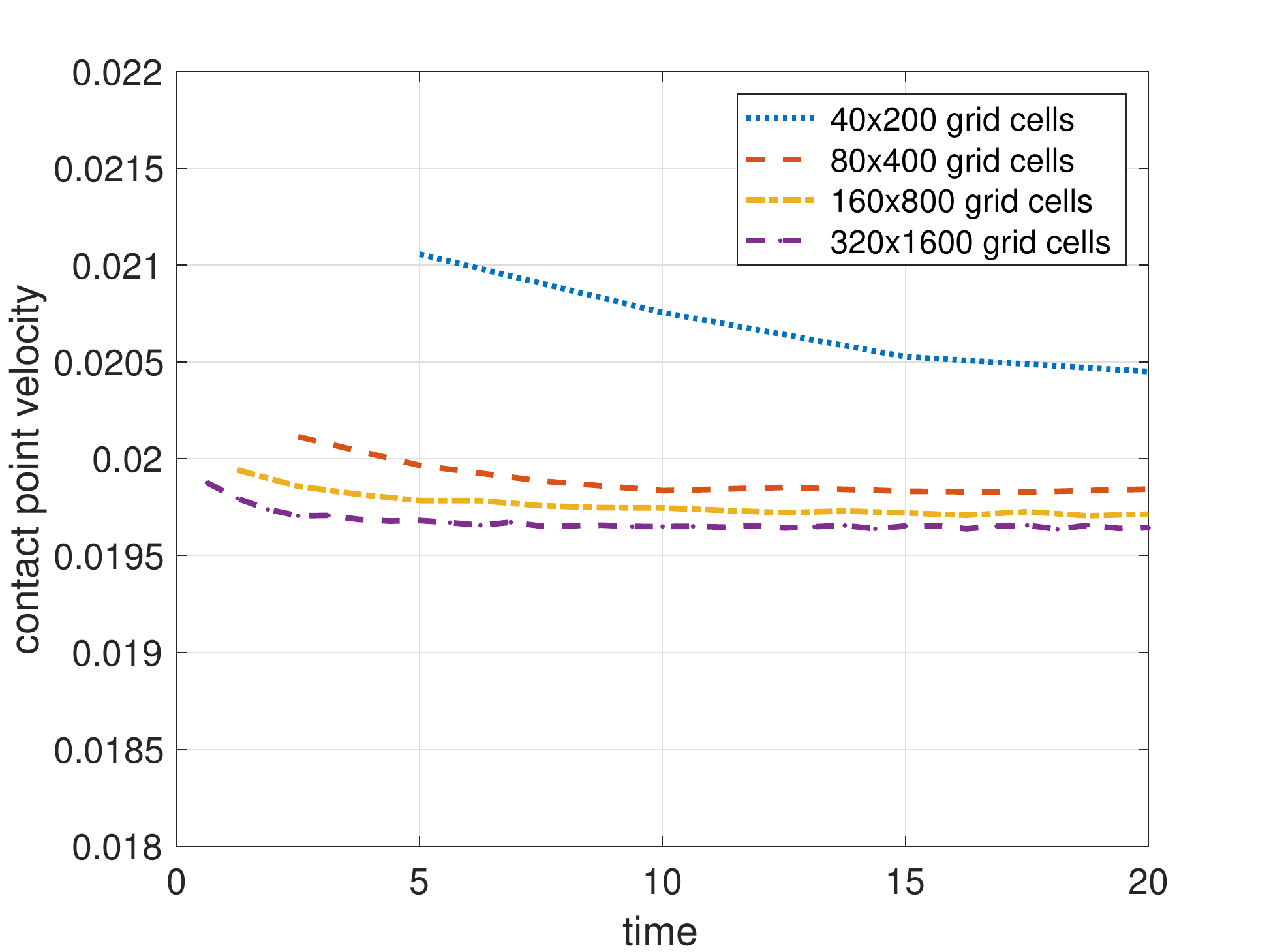}
    \caption{\figtext{Convergence study for the Preliminary Test.}}
    \label{fig:test1}
\end{figure}

\Figref{fig:test1} shows resulting contact point velocities as a function of time (average over 25 time steps) for the different meshes.
The error in the average computed velocity for the finest mesh over the time span $t=[15,20]$ is $1.8\%$.  
We see that the numerical solutions converge as the computational mesh is refined. However, the convergence is to a solution with a slightly different speed than the prescribed. Possible reasons for the  approximately $1.8\%$ discrepancy of the limiting solution will be discussed below.

There are grid related oscillations in the contact point velocity, similar to those seen in \textbf{Figures \ref{fig:analytic}\hspace{-0.2cm}--\ref{fig:compare_deltas}}, but their amplitude decrease with mesh refinement for the initial refinements. At a refinement level of $h=0.0125$  the decay of the oscillations seems to stagnate. We believe the remaining oscillations are triggered by similar processes as the spurious velocities observed in  \cite{zahedispurious}, and references therein.

\newpage
\subsection{Channel Flow}
\label{sec:Test2}
In this section we present results for the full problem of capillary fluid-fluid displacement in a horizontal channel. Channels of non-dimensional lengths $L=120,\,240$ and $480$ are used, all with a width of $W=20$. Non-dimensional parameters are set to $\mu_1=0.3$, $\mu_2=1$, $\rho_1=1$, $\rho_2=0.73$, $\sigma = 1$, which corresponds to oil being displaced by water. For this set-up the static contact angle is $140$ degrees measured from the oil side.

\begin{figure}[h!]
  \centering
    \includegraphics [width=0.5\columnwidth]{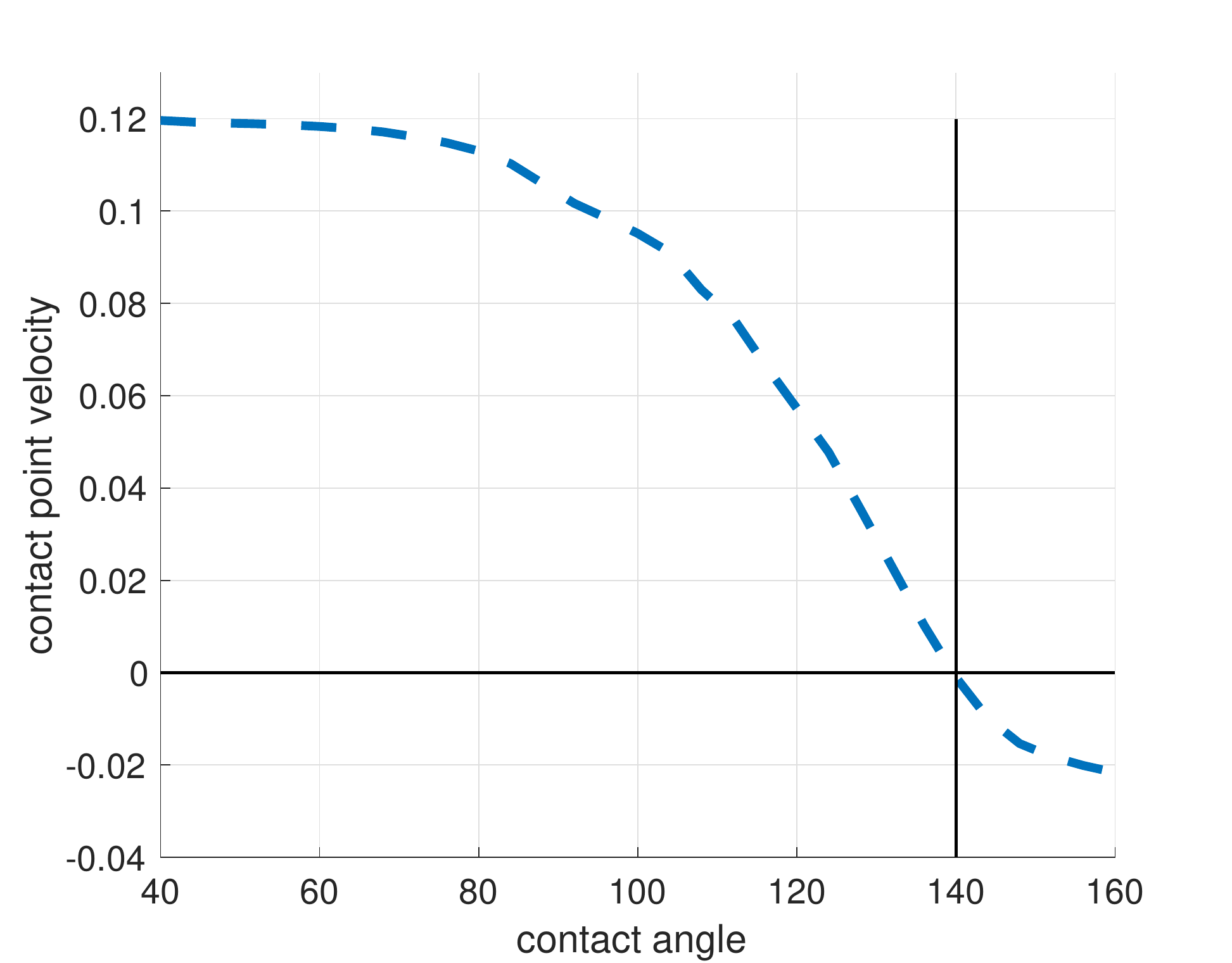}
    \caption{\figtext{Micro model result: Tabulated relation between non-dimension contact point velocity $U$ and contact angle $\phi$.}}
    \label{fig:micro_vel}
\end{figure}

The micro model is first used to pre-compute relations between the contact point velocity and contact angle. The set-up of the micro model is the same as in \cite{MartinMicro} and the results are the ones for a microbox of non-dimensional size $30$, which is much larger then the non-dimensional diffusion length of $1$ (see \cite{MartinMicro}). The microdomain size is sufficient to include most of the viscous bending, see the discussion in \cite{MartinMicro} for more details. The resulting relation between contact angles and velocities obtained from the microsimulations is plotted in \Figref{fig:micro_vel}. Note that to get the corresponding dimensional velocities one would need to take into account the scaling by the reference velocity $U_{ref}=\sigma/\mu$. However, we have used the same scalings for both the macro and micro models, and can therefore directly read the non-dimensional values from \figref{fig:micro_vel}.

Now, macroscopic simulations are preformed using the tabulated data in \figref{fig:micro_vel}. In each time step the velocity boundary conditions developed in \secref{sec:macroscopicboundaryconditions} are applied according to the given contact point velocities from the micro model, depending on the calculated angle. The height and width of the bump function (\secref{sec:artificialboundary}) are here $\delta=0.5$ and $w=4 \delta$ (i.e.\ $\delta=0.025W$). We assume  the excluded part of the domain has  a size that  corresponds to the scale of the intermediate region according to the discussion in \secref{sec:matching}.

To begin with we consider the shortest channel and  run the simulations for a total time of $T=2000$ ($T=800$ for the finest mesh) using three different meshes with $40\times240$, $80\times480$ and $160\times960$ grid cells, respectively.  Corresponding mesh sizes are $h=0.5$, $h=0.25$ and $h=0.125$. 
 Time steps of $\Delta t=2$, $\Delta t=1$ and $\Delta t=0.5$ are used. The apparent contact angle is calculated at a height of $D=0.5$ for all meshes (i.e.\ at the height of one coarse grid cell above the wall). 

\begin{figure}[h!]
  \centering
    \includegraphics [width=0.65\columnwidth]{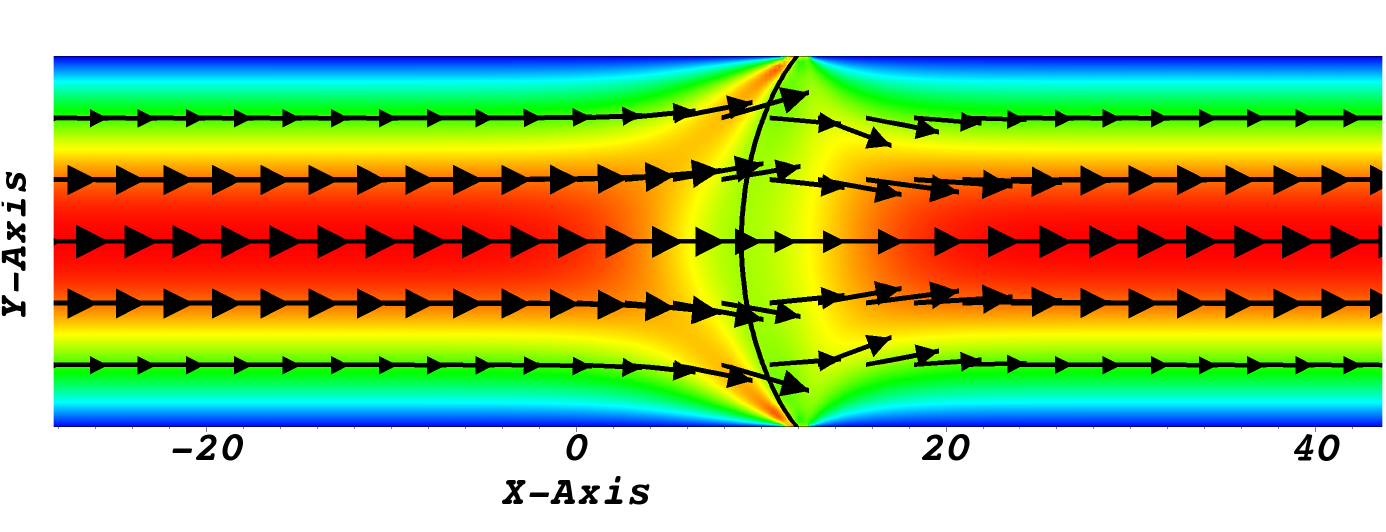} 
    \caption{\figtext{Velocity field and fluid interface at the end of the simulation for channel length L=120, mesh size h=0.25. Note that only very few velocity vectors are presented, compared to the degrees of freedom in the simulation. }}
    \label{fig:test3_vel_field}
\end{figure}

For all simulations the initial interface is a vertical line located at a non-dimensional distance of 25 into the channel (measured from the left end) and the initial velocity is zero in the whole domain. After an initial transient the system goes into a quasi-steady state, where the flow is determined by a balance of capillary forces and viscous stress. At the quasi-steady state the solution consists of a curved interface, and a Poiseuille flow profile away from the interface, see \figref{fig:test3_vel_field}.

\begin{figure}[h!]
    \centering
    \subfloat[\textbf{a.}\ \ $L=120$ \label{fig:L120analytic}]{%
       \includegraphics[width=0.5\columnwidth]{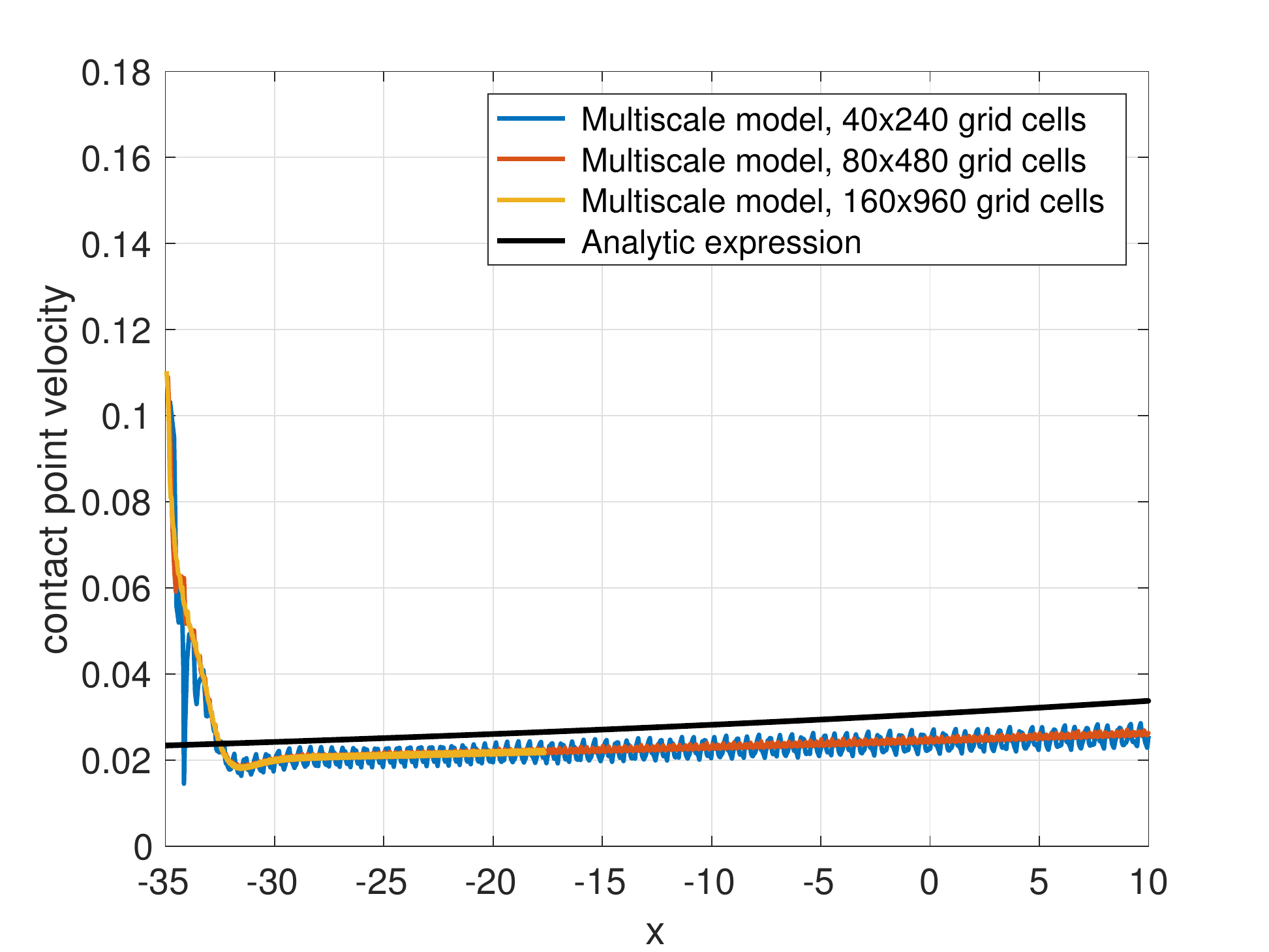}
     }
          \subfloat[\textbf{b.}\ \ L=240 \label{fig:L240analytic}]{%
       \includegraphics[width=0.5\columnwidth]{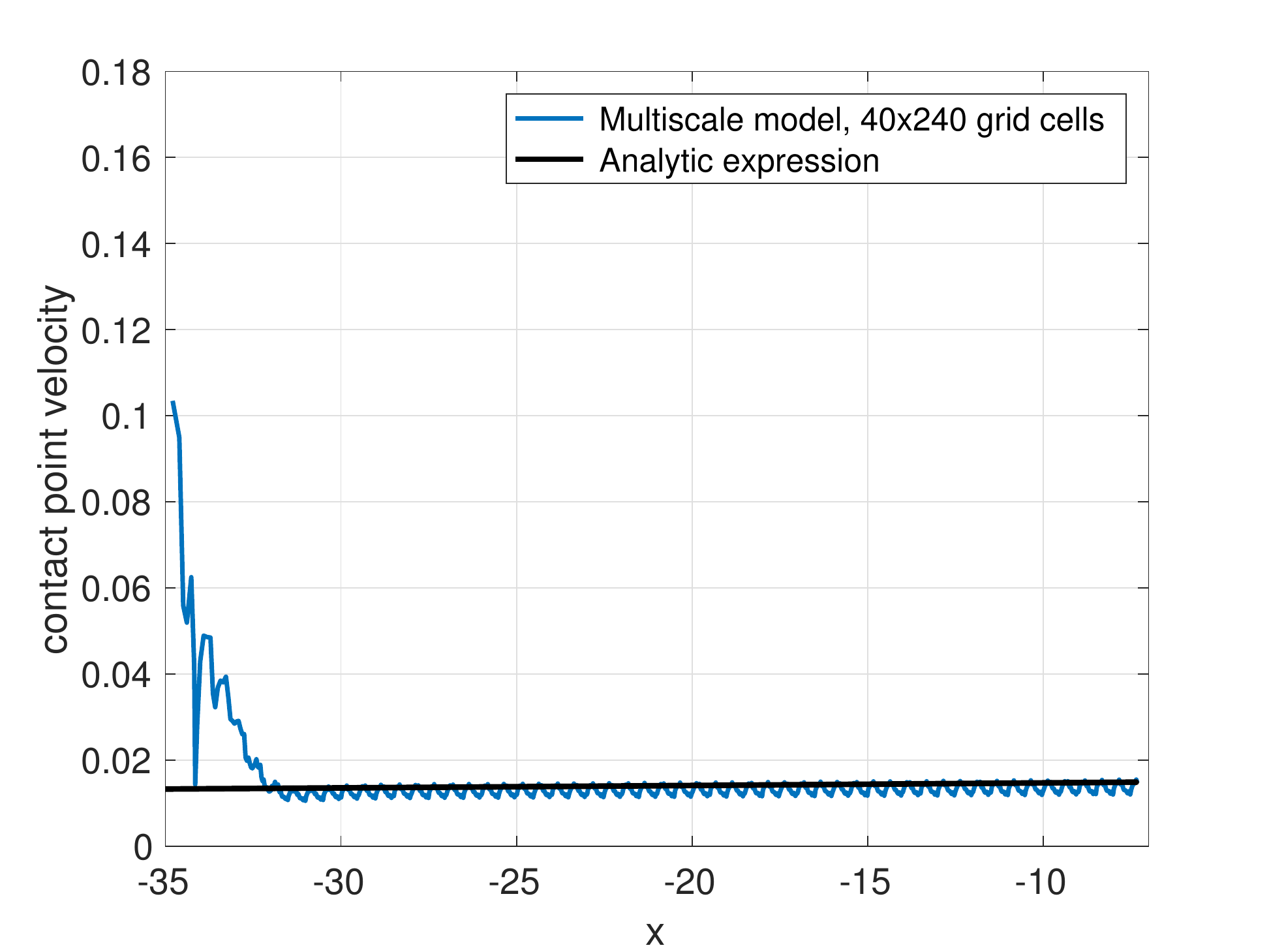}
     }
     
              \subfloat[\textbf{c.}\ \ L=480 \label{fig:L480analytic}]{%
       \includegraphics[width=0.5\columnwidth]{CP_vels_Test2_L240_analytic}
     }
    \caption{\figtext{Contact point velocity $U$ as a function of the contact point position $x$. The results are compared to the analytic expression in \eqref{eq:analytic}, which is valid for the limiting shape of a perfect circular arc.}}
    \label{fig:analytic}
\end{figure}

In \figref{fig:L120analytic} the contact point velocity is plotted as a function of time. In Table \ref{tab: channel convergence} we compare local  time averages of contact point velocities, and amplitudes of grid-level oscillations, for the three meshes.  The amplitude is computed from maximum and minimum values in the time interval $500\le t\le520$. The convergence rate for the time-averaged velocity is over $3$, but there is a  stagnation in convergence for the grid-level oscillations also in these computations.

\begin{table}[h!]
\begin{center}
\begin{tabular}{c|ccc}
h & Mean velocity & Convergence order & Amplitude \\
& & (mean velocity) &\\
\hline
0.5&0.0209308&&0.0022026\\
0.25&0.0212261&&0.0005789\\
0.125&0.0211924&3.135&0.0004236
\end{tabular}
\end{center}
\caption{Convergence of contact point velocity. Mean velocity, and amplitude of grid-level oscillations, for  the $L=120$ channel when $500\le t\le520$.}
\label{tab: channel convergence}
\end{table}

We must expect a discrepancy for converged solutions, similar to the one discussed above for the Preliminary Test, also in this case. For this test there is no exact solution to compare with, but as the length of the channel increases the flow is expected to approach a limiting state with an interface shaped of a perfect circular arc. There is an analytic formula describing the movement of such an interface. In \cite{exactvel} a theoretical expression for the interface velocity for a three-dimensional fluid-fluid displacement in a horizontal capillary channel is derived. We have modified the expression to be applicable for two-dimensional flow (equivalent to three-dimensional flow between two infinite parallell plates) and get
\begin{equation}
v(t)= \frac{W \sigma \cos (\pi-\phi_s)}{6\mu_2\left[L-x_{cp}(t)\left(1-\frac{\mu_1}{\mu_2}\right)\right]} ,
\label{eq:analytic}
\end{equation}
where $x_{cp}$ is the distance between the contact point position and the left end of the channel and $\phi_s$ is the static contact angle.

In \figref{fig:analytic} the simulated contact point velocities are shown for the three channel lengths, together with the corresponding limiting velocities \eqref{eq:analytic}. The acceleration of the contact point is due to that the more viscous fluid 2 (oil) is displaced by the less viscous fluid 1 (water). 
For the two longer channels the multiscale simulation results agree well with the theoretical expression for the interface velocity. For the shorter channel, $L=120$, the  simulation predicts a lower velocity compared to the velocity in the limiting case, given by \eqref{eq:analytic}.

\begin{figure}[h!]
  \centering
    \includegraphics [width=0.5\columnwidth]{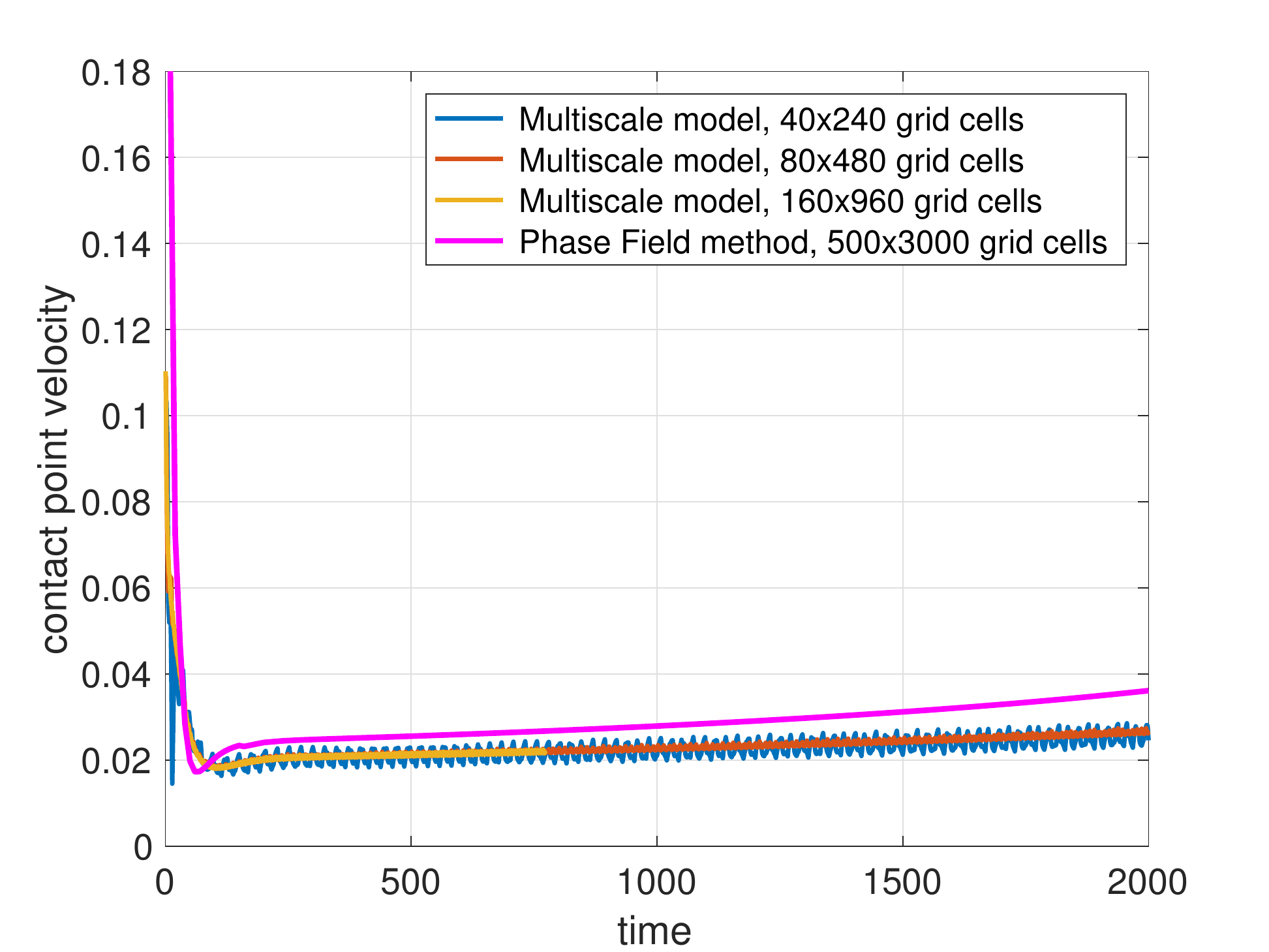} 
    \caption{\figtext{Contact point velocity $U$ as a function of time. The results are compared to a phase field simulation.}}
    \label{fig:L120phasefield}
\end{figure}

Another comparison is also possible.
In \cite{MartinMicro} phase field simulations of capillary fluid displacement with a set up equal to ours are presented. In \figref{fig:L120phasefield} we compare our results for the shorter channel with the result from the full phase field simulation in \cite{MartinMicro}. Again, the multiscale model predicts a lower contact point velocity. In fact, the phase field simulation predicts a velocity in good agreement with \eqref{eq:analytic}. 

\subsection{ Sensitivity of the method}
Above we have demonstrated that with our model for contact point dynamics and our implementation thereof, numerical solutions converge as the computational mesh is refined, to a solution in the vicinity of the correct solution, but with a small discrepancy. We identify two possible reasons for the discrepancy. One possibility is that the interface is not completely flat at the matching region. This has two consequences: the Huh and Scriven solution is only an approximation and it becomes difficult to define and calculate the apparent contact angle. Another possibility is errors due to the implementation strategy of the velocity boundary conditions: the Huh and Scriven velocity is read along a contour (to avoid the singularity precicely at the contact point) but imposed as a velocity boundary condition at the solid wall, and not along the corresponding contour (see \secref{sec:artificialboundary}).

To understand the sensitivity to the specific  choices in our implementation we have made some variations in the implementation. In particular, we changed the contour determining the velocity boundary condition, and how the apparent angle was determined. Results from these  numerical tests are reported below. 

\subsubsection{Effect of angle calculation}
\begin{figure}[h!]
    \centering
           \subfloat[Velocity as a function of position, compared to limiting velocity in \eqref{eq:analytic}.  \label{fig:L120phasefield_angleAt10}]{%
       \includegraphics[width=0.5\columnwidth]{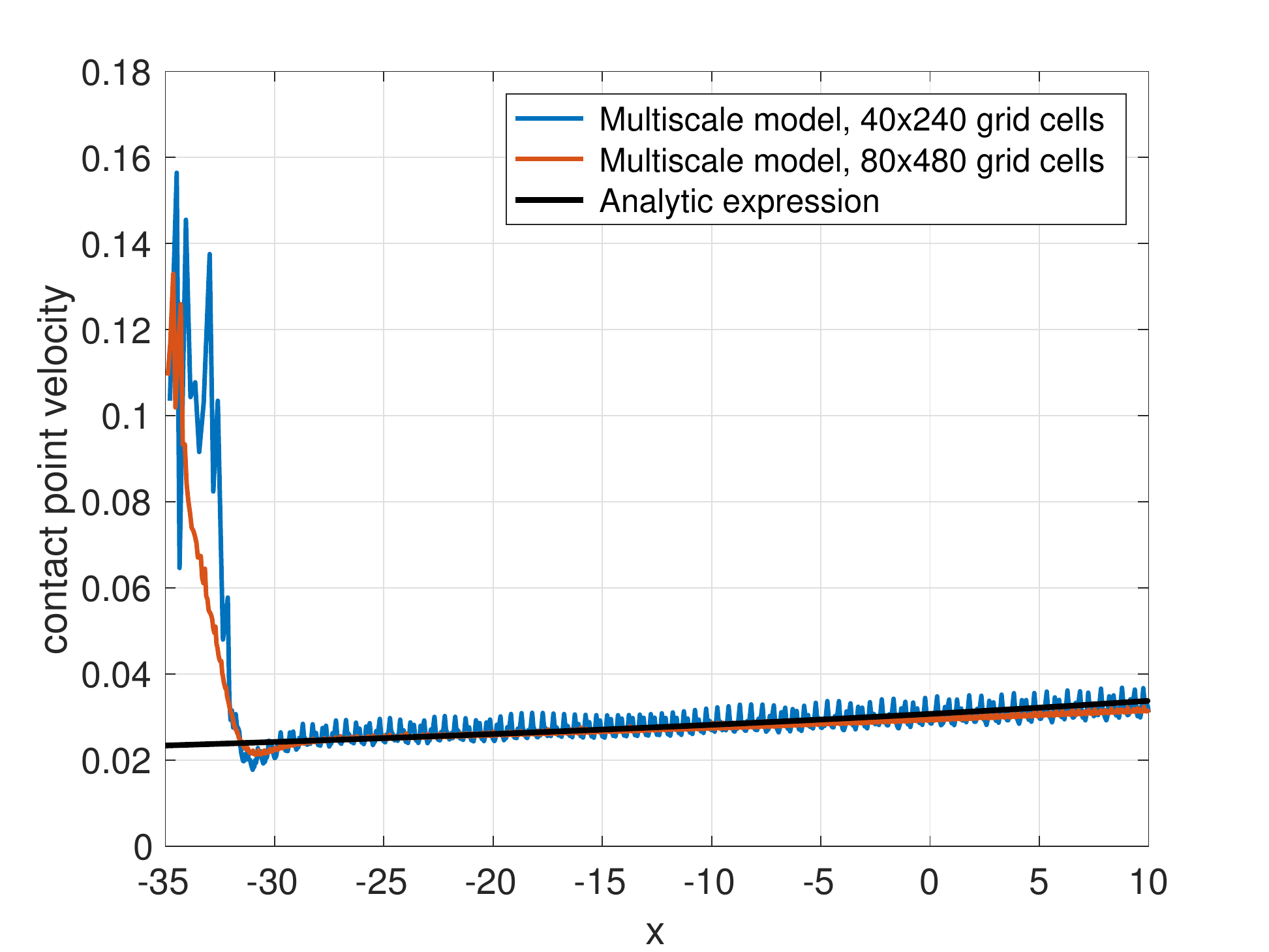}
       }
           \subfloat[Velocity as a function of time, compared to phase field simulation.  \label{fig:L120phasefield_angleAt10}]{%
       \includegraphics[width=0.5\columnwidth]{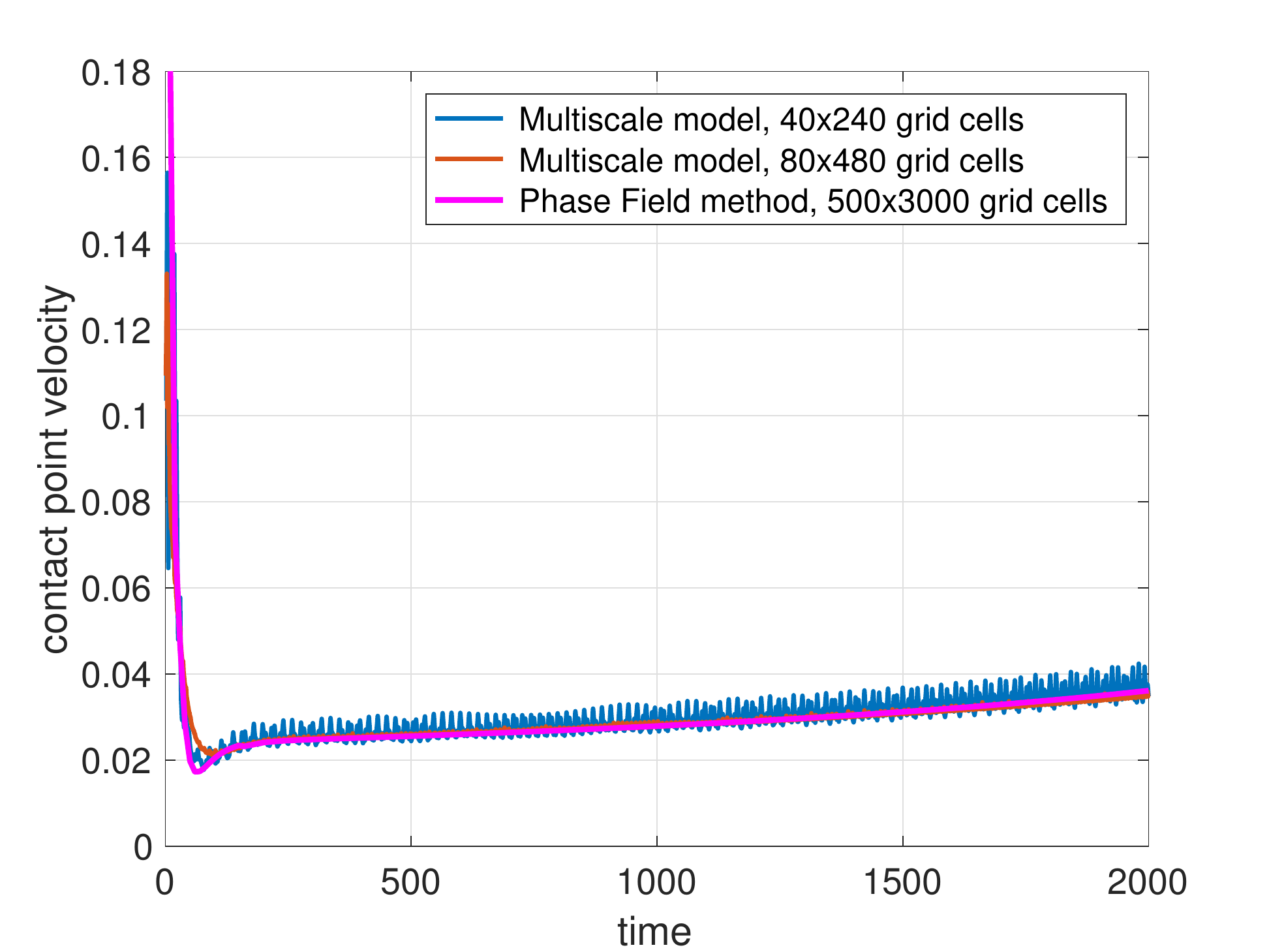}    
       }
    
    \caption{\figtext{Contact point velocity for the $L=120$ case when the apparent angle is calculated with $D=10$.}}
    \label{fig:angleAt10}
\end{figure}

From the phase field solutions in \cite{MartinMicro} it is observed that the interface shape is essentially circular in the central part of the domain throughout the simulations. However, it is also observed that the interface is bent close to the contact point, and this effect is larger for the shorter channels. In fact, the relation between the apparent angle and the interface speed in the phase field simulation of channel flow, is reported to be consistent with the results from the microsimulations, only when the apparent angle is computed from the interface shape at a certain distance (non-dimensionalized by  diffusion length) from the wall. 
We also see a corresponding effect here when comparing angle calculations at different heights $D$ above the wall (see \secref{sec:angle} for definition of $D$).  For the longer channels we observe that angles calculated at different heights agree well with each other, but for the shorter channel there is a difference in the calculated angle depending on the height $D$. With this motivation we perform a simulation for the shorter channel , $L=120$, where the angle is calculated at a height $D=10$ (i.e.\ at the center of the channel) instead of at the height $D=0.5$ used above. The result is presented in \figref{fig:angleAt10} and we see that the resulting contact point velocities now agree well both with the theoretical expression \eqref{eq:analytic} and the phase field solution. We also see that with grid refinement the result is converging towards the theoretical result/the phase field result, and that the oscillations decrease.  We also note that in the present simulation the number of grid points is only a small fraction as compared with the phase field simulation. 

\subsubsection{Implementation of the velocity dirichlet condition}
\begin{figure}[h!]
    \centering
               \subfloat[\label{fig:compare_deltas_1}]{%
       \includegraphics[width=0.5\columnwidth]{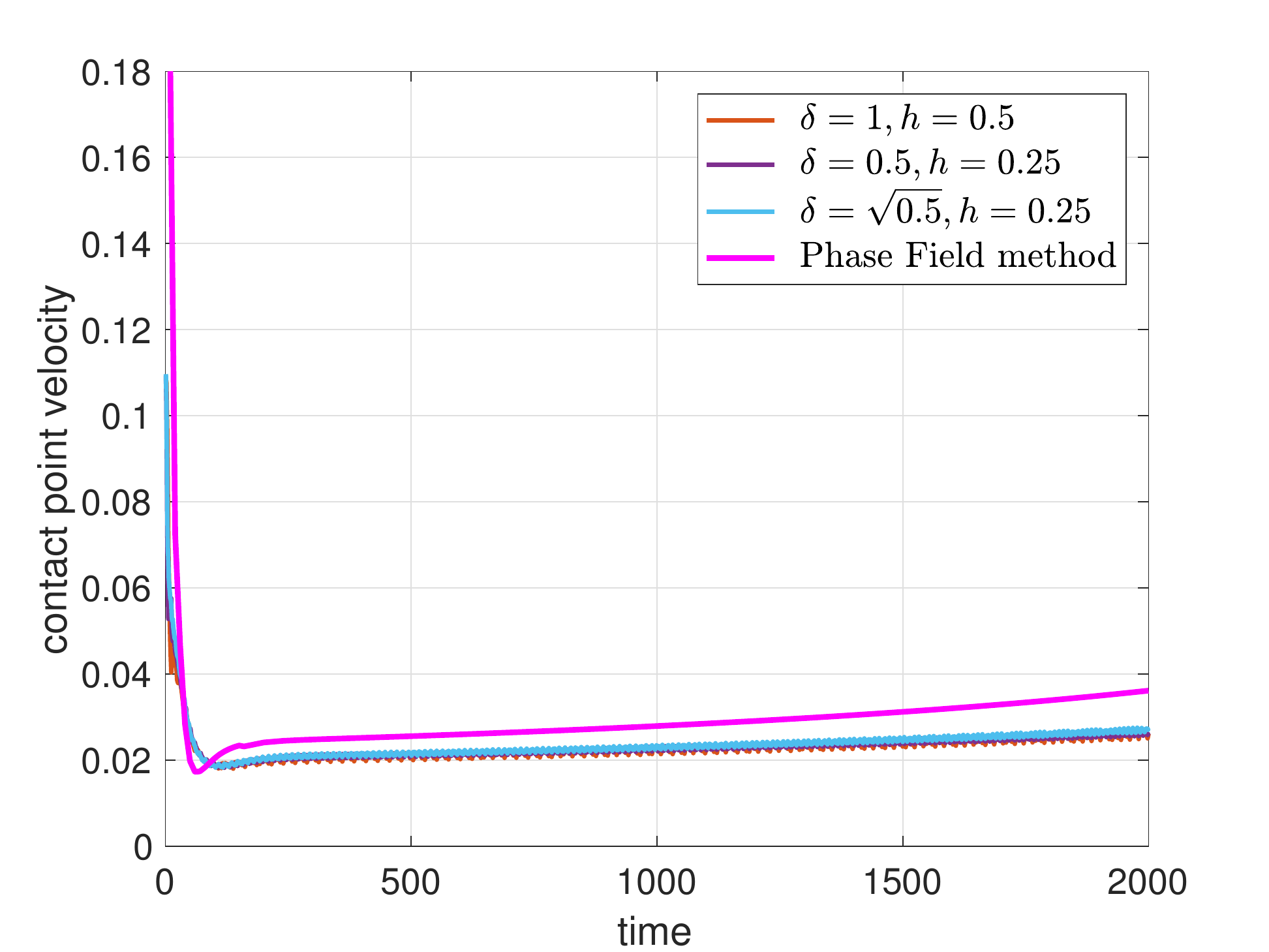}    
       }
                  \subfloat[\label{fig:compare_deltas_zoom}]{%
       \includegraphics[width=0.5\columnwidth]{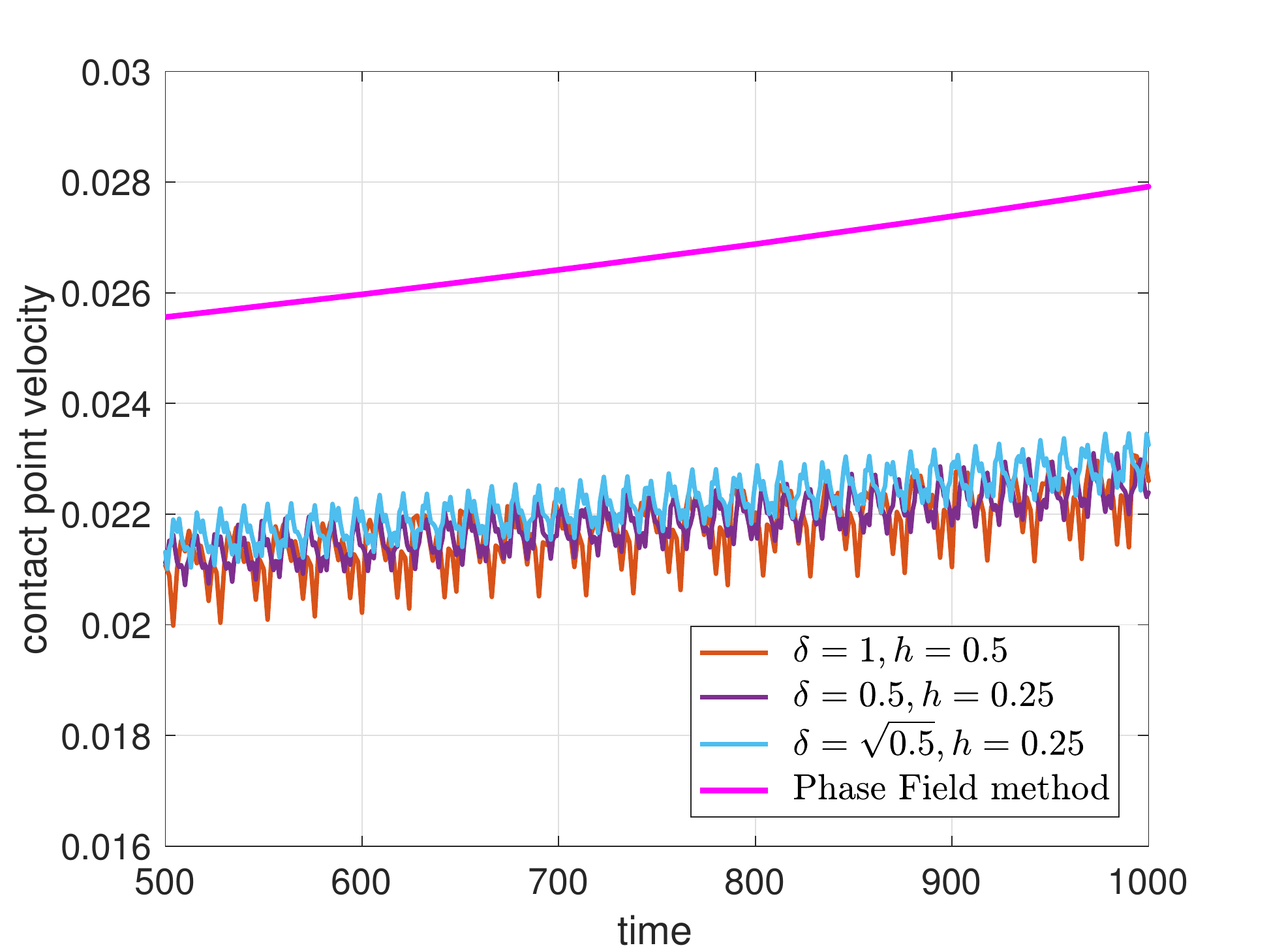}    
       } 
     ~
    \caption{\figtext{Investigating the effect of $\delta$ and $w=4\delta$ for different resolutions $h$. The right figure is a zoomed version of the left figure.}}
    \label{fig:compare_deltas}
\end{figure}

The aim of this section is to investigate the effect of the modified boundary in \secref{sec:macroscopicboundaryconditions} and the simple approach of applying the boundary conditions (described in \secref{sec:artificialboundary}). 
Simulations are performed for the shorter channel $L=120$, with different sizes of the excluded part of the domain, i.e.\ different values of $\delta$ (and $w=4\delta$). We also vary the spatial mesh size for different $\delta$, to investigate the effect of resolving the small scale features of the peak in the velocity boundary function (\secref{sec:artificialboundary}). In \figref{fig:compare_deltas} we present results for three different sets of $\delta$ and mesh sizes $h$. 

For the simulation where $\delta=0.5, h=0.25$ the size of the excluded part is half the size compared to the simulation where $\delta=1, h=0.5$, but the resolution of the boundary function is the same (i.e.\ the same number of grid points to resolve the peak in the boundary velocity function). Comparing the two simulations we see that the result slightly improve for the case with smaller $\delta$. The effect is however small compared to the effect of changing the way the contact angle is estimated, compare to \figref{fig:angleAt10}. To investigate the effect of resolving the small scale features of the boundary function, the result for the simulation with $\delta=0.5, h=0.25$ is compared to a simulations where $\delta=\sqrt{0.5}, h=0.25$, i.e.\ larger $\delta$ but also a higher resolution. In \figref{fig:compare_deltas} we see that a larger $\delta$ with higher resolution only slightly improves the result as well. This indicates the error due to the simple approach of applying the boundary condition (\secref{sec:artificialboundary}) does not have an significant effect on the result.

\section{Conclusions and Discussion}
\label{sec:conclusions}

We have presented a new idea for including contact point dynamics in standard two-phase models. The idea is based on multi-scale modeling, where the result from a local micro model for contact point dynamics (presented in  \cite{MartinMicro}) is used. The micro model  includes nano-scale physics, such as molecular diffusion, which is not present  in  the standard Navier--Stokes model for two-phase flow. The effect of the micro model  is communicated to the global model via special macroscopic boundary conditions. The boundary conditions are based on the Huh and Scriven analytic velocity for steady movement of a contact point \cite{HUH}. 

The approach is very general, and can be modified to include other nano-scale effects that may be important for the large scale dynamics.
The so called contact-line friction is an example, see \cite{Amberg}.  This effect can be incorporated in our method straightforwardly by changing the boundary condition in the micro problem. 

The presented method is exemplified in computations of $2D$  fluid-fluid displacement in a channel. The potential of our technique is demonstrated by the fact that we can achieve as accurate result as a full phase field simulation at a fraction of the computational cost. Accuracy for a phase field simulation  relies on  resolving features at   the length and time scales of molecular diffusion, while for our method such resolution is not required. Our computations demonstrate convergence  as the computational mesh is refined. A  correct solution is known in a few simplified cases, and then the converged solution is in the vicinity of the correct solution (a  $1.8\%$ discrepancy was observed).  We discuss reasons for the discrepancy, and have found that results are sensitive to the precise algorithm for estimating  the apparent contact angle, while less so to several other implementational choices. 

For a realistic channel flow model we compare velocities from our simulations  with the theory for a limiting steady interface displacement with an interface shape of a perfect circular arc. 
For  the longer channels, the resulting velocities show good agreement.  It is reasonable that the results for the longer channels show better agreement, since the flow and interface shape is approaching the limiting case as the channel length increase.
We have found that  for the shorter channel our results are more sensitive to how the contact angle is estimated, then for the longer channels. This shows that further studies on how to estimate the apparent angle are needed.

However, we believe there are possible gains in also improving the implementations of the velocity boundary condition, the level set boundary conditions, the reinitialization procedure and the curvature calculations, especially in the vicinity of the contact points. 
For example, an ideal reinitialization would preserve both contact point positions and angles between the interface and the boundary.  Possibly a geometric reinitialization would be able to fulfill  the requirements better than our approach does, however probably at a cost of efficiency. Similarly, the advection step should move the contact point to a new position, while the angle evolves according to the local velocity. We have tried a few different combinations of boundary conditions, but found that applying a Dirichlet condition in the advection of the level set function leads to a distortion of the contact angle. More research in this area is likely to improve the accuracy of the methodology.

Another area where improvements are needed is related to  the calculation of the interface curvature. Here, the calculation is carried out by projecting the divergence of the normals (equation \eqref{eq:normcurv1}), including a diffusion term of size $4h^2$, onto the space of linear elements, see \secref{sec:discimp} and \cite{MartinHPC}. This means solving an elliptic partial differential equation with a laplace-term. With no contact points present a homogenous Neumann boundary condition suffices, while such a condition leads to numerical artefacts in the calculated curvature close to contact points. As discussed in \secref{sec:discimp} we have no boundary conditions for the calculation of the curvature. An area for further research  could be to find suitable boundary conditions for the curvature calculation.  

Further improvement is possible by applying the macroscopic velocity boundary condition at the modified boundary, instead of the simple approach used here, and described in \secref{sec:discimp}. One possible way forward would be to use some kind of immersed methodology, for example  cut-FEM. A starting point could be the cut-FEM method for handelling a fluid  interface described in \cite{ZahediStokesInterface}.

\bibliography{references}

\end{document}